\documentclass[aps,reprint,superscriptaddress,amsmath,citeautoscript,longbibliography]{revtex4-1}
\usepackage{graphicx}

\usepackage[breaklinks=true]{hyperref}
\usepackage[all]{hypcap}
\hypersetup{
	colorlinks = true,
	linkcolor = {blue},
	citecolor = {blue},
	urlcolor = {blue},
}

\usepackage{multirow}

\usepackage{tabularx}
\newcolumntype{Y}{>{\centering\arraybackslash}X}

\AtBeginDocument{%
    \newwrite\bibnotes
    \def\bibnotesext{Notes.bib}
    \immediate\openout\bibnotes=\jobname\bibnotesext
    \immediate\write\bibnotes{@CONTROL{REVTEX41Control}}
    \immediate\write\bibnotes{@CONTROL{%
    apsrev41Control,author="08",editor="1",pages="1",title="0",year="1"}}
     \if@filesw
     \immediate\write\@auxout{\string\citation{apsrev41Control}}%
    \fi
}%

\begin{document}

% Title and authors
\title{Diffuse scattering from dynamically compressed single-crystal zirconium following the pressure-induced $\alpha\to\omega$ phase transition}
\author{P.~G.~Heighway}
\email{patrick.heighway@physics.ox.ac.uk}
\affiliation{Department of Physics, Clarendon Laboratory, University of Oxford, Oxford, UK}
\author{S.~Singh}
\affiliation{Lawrence Livermore National Laboratory, Livermore, CA, USA}
\author{M.~G.~Gorman}
\affiliation{Lawrence Livermore National Laboratory, Livermore, CA, USA}
\author{D.~McGonegle}
\affiliation{AWE, Aldermaston, Reading, Berkshire, UK}
\author{J.~H.~Eggert}
\affiliation{Lawrence Livermore National Laboratory, Livermore, CA, USA}
\author{R.~F.~Smith}
\affiliation{Lawrence Livermore National Laboratory, Livermore, CA, USA}
\date{\today}

%%  0. Abstract
\begin{abstract}
The prototypical $\alpha\to\omega$ phase transition in zirconium is an ideal test-bed for our understanding of polymorphism under extreme loading conditions. After half a century of study, a consensus had emerged that the transition is realized via one of two distinct displacive mechanisms, depending on the nature of the compression path. However, recent dynamic-compression experiments equipped with \textit{in situ} diffraction diagnostics performed in the past few years have revealed new transition mechanisms, demonstrating that our understanding of the underlying atomistic dynamics and transition kinetics is in fact far from complete. We present classical molecular dynamics simulations of the $\alpha\to\omega$ phase transition in single-crystal zirconium shock-compressed along the [0001] axis using a machine-learning-class potential. The transition is predicted to proceed primarily via a modified version of the two-stage Usikov-Zilberstein mechanism, whereby the high-pressure $\omega$-phase heterogeneously nucleates at boundaries between grains of an intermediate $\beta$-phase. We further observe the fomentation of atomistic disorder at the junctions between $\beta$ grains, leading to the formation of highly defective interstitial material between the $\omega$ grains. We directly compare synthetic x-ray diffraction patterns generated from our simulations with those obtained using femtosecond diffraction in recent dynamic-compression experiments, and show that the simulations produce the same unique, anisotropic diffuse scattering signal unlike any previously seen from an elemental metal. Our simulations suggest that the diffuse signal arises from a combination of thermal diffuse scattering, nanoparticle-like scattering from residual kinetically stabilized $\alpha$ and $\beta$ grains, and scattering from interstitial defective structures.
\end{abstract}
\maketitle

%%  1. Introduction
\section{\label{sec:introduction} Introduction}

When dynamically compressed to planetary pressures and temperatures, even the most apparently simple elements can exhibit rich allotropy. Among the best-known examples are: the Group-XI metals Cu, Ag, and Au, all of which undergo a face-centered cubic (fcc) to body-centered cubic (bcc) phase transition at just under 200~GPa \cite{Sharma2020b,Sharma2020c,Sharma2019,Briggs2019}; the Group-IV metals Ti, Zr, and Hf, which follow a hexagonal close-packed (hcp) to simple hexagonal (hex-3) to bcc transition path over a few tens of gigapascals \cite{Xia1990,Xia1990b}; the Group-XIV semiconductor Si, which experiences three phase transitions along the Hugoniot before shock-melting at just 27~GPa \cite{Jamieson1963,McMahon1993,Olijnyk1984,McBride2019}; and its congener C, whose elusive BC8 `superdiamond' phase transition at 1~TPa remains the subject of intense investigation \cite{Lazicki2021,Shi2023,NguyenCong2024}. With the worldwide proliferation of high-power, long-pulse laser-compression platforms that allow us to shock- or ramp-compress solid targets rapidly and reproducibly to gigapascal-scale pressure states -- coupled to ultrafast x-ray diagnostics that capture transformations of their crystal structure \textit{in situ} within the nanosecond window before they disintegrate -- the experimental observation of polymorphism under extreme loading conditions has become almost routine. The study of pressure-induced polymorphism informs our understanding of condensed matter physics at its most fundamental level, and at a rate that will only increase in the coming decade with the arrival of high-repetition-rate instruments such as the DiPOLE 100-X laser at the European X-ray Free-Electron Laser (EuXFEL) \cite{Gorman2024}.

With the high-pressure phase diagrams of many important elements now firmly established (along the Hugoniot, at least), the focus of current phase-transition studies is shifting towards the identification of the atomistic mechanisms by which transformations are realized. Displacive solid-solid phase transitions are characterized by a unique atomistic pathway, each of which brings about a high-pressure daughter phase with a particular orientation relationship (OR) with the ambient parent phase. The orientation of the daughter phase is readily deduced from the azimuthal angles at which its diffraction peaks do or do not appear, while the orientation of the parent phase is often known in advance by virtue of the target being a single crystal. Experimentally determined ORs have successfully been used to discriminate between rivaling atomistic transformation mechanisms in Fe \cite{Hawreliak2006}, Si \cite{Pandolfi2022}, Cu \cite{Sims2022}, and Zr \cite{Singh2024}. Studies offering this depth of analysis are currently rare, but are absolutely essential to developing our understanding of the lattice-level dynamics associated with solid-solid phase transitions.

The nature of the $\alpha\to\omega$ (hcp$\,\to\,$hex-3) transition in zirconium is arguably the most complex and contentious of any elemental metal. Decades of experimental study have revealed that the ductile-to-brittle $\alpha\to\omega$ phase transition is generally realized via one of two inequivalent mechanisms. The first yields an $\omega$-phase that assumes the so-called TAO-I OR with the parent $\alpha$ crystal, and is usually observed in samples recovered from uniaxial shock-compression scenarios \cite{Song1995,Jyoti1997,jyoti2008}. The second mechanism produces an $\omega$-phase that takes the Silcock OR, and is more often observed during static compression or under high-pressure torsion \cite{Silcock1958,Sargent1971,Rabinkin1981,Wenk2013,Adachi2015,Wang2019}. These trends do not hold universally, however: in several instances the `converse' mechanism has been observed \cite{UZ1973,Kutsar1990}, and, in a dynamic-compression study by Swinburne \textit{et al.}, the authors reported an entirely new OR \cite{Swinburne2016} (the interpretation of which was questioned soon thereafter in Ref.~\cite{Armstrong2021}). The situation became yet more complicated following a very recent experimental-computational study by Singh \textit{et al.}\ \cite{Singh2024}. Using femtosecond x-ray diffraction to characterize the $\alpha\to\omega$ phase transition in single-crystal Zr shock-compressed along its [0001] direction, the authors measured three distinct ORs \emph{simultaneously} (namely the TAO-I, the Silcock, and another previously unreported OR), demonstrating that the $\alpha\to\omega$ transition was being realized via three competing, concurrent mechanisms. No single account of the phase-transition kinetics in Zr yet exists that harmonizes these seemingly disparate results; clearly, there is still a great deal to be learned about (and from) high-pressure polymorphism in zirconium.

In addition to the multiple competing $\alpha\to\omega$ transition mechanisms, Singh \textit{et al.}\ observed a second curious feature in the data: a diffuse scattering signal of comparable total intensity to the Bragg peaks that exhibited a pronounced, sixfold-symmetric structure around the beam direction. The diffuse signal would appear to suggest atomistic disordering of some description, but its azimuthal structure rules out a simple isotropic liquid or glass. In that previous study \cite{Singh2024}, the authors briefly presented the results of classical molecular dynamics (MD) simulations of shock-compressed, [0001]-oriented Zr single crystals that reproduced the same diffuse, azimuthally structured diffraction signal, but did not discuss its source. The purpose of the present study is to describe the aforementioned MD simulation campaign in complete detail. We will examine the rich and complex microstructural evolution precipitated by the $\alpha\to\omega$ phase transition, focusing on the displacive and diffusive mechanisms through which the transition is realized. We will then interrogate the reciprocal-space structure of the shock-loaded system, and make direct comparisons between experimental and synthetic diffraction signals. By combining these real- and reciprocal-space analyses, we will explain the origin of the unique diffuse scattering signal measured experimentally by Singh \textit{et al.} \cite{Singh2024}.

The paper is laid out as follows. We first give a detailed description of our simulation methodology in Sec.~\ref{sec:methodology}, including details of the interatomic potential used, the shock-compression technique, and the characterization techniques employed. The results of the simulations are given in Sec.~\ref{sec:results}, which is subdivided into real-space (Sec.~\ref{subsec:real-space}) and reciprocal-space (Sec.~\ref{subsec:reciprocal-space}) analyses. We then discuss the implications of our results in Sec.~\ref{sec:discussion} before concluding in Sec.~\ref{sec:conclusion}.

% Figure 1: High-pressure phase diagram and Hugoniot of zirconium
\begin{figure}[b]
\centering
\includegraphics{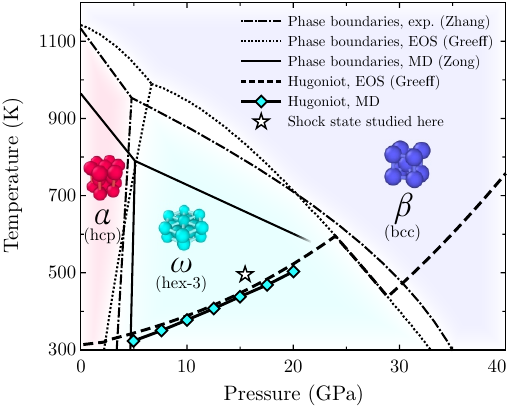}
\caption{\label{fig:phase_diagram} Hydrostatic high-pressure phase diagram and principal Hugoniot of zirconium. Shown are phase boundaries between the $\alpha$ (hcp), $\beta$ (bcc), and $\omega$ (hex-3) phases determined in cubic-anvil-based experiments by Zhang \textit{et al.}~\cite{Zhang2005}, from a three-phase, semi-empirical equation of state (EOS) by Greeff~\cite{Greeff2005}, and from static molecular dynamics simulations conducted with the Zong potential~\cite{Zong2018}. Also shown is the Hugoniot from the Greeff EOS and from the Zong potential, the latter being approximated by solving the Rankine-Hugoniot equation $E-E_0=\frac{1}{2}P(V_0 - V)$ for a perfect $\omega$ single crystal. The shock state examined here is marked by a star.}
\end{figure}

%%  2. Methodology
\section{\label{sec:methodology} Methodology}
%Here, we describe the design and implementation of our classical molecular dynamics simulations, which, due to the nature of both the interatomic potential and the shock-loading geometry that they use, require a relatively elaborate explanation.

%  2.1 Interatomic potential
\subsection{\label{subsec:potential} Interatomic potential}

The choice of the classical potential governing the microscopic forces exerted between atoms -- and from which all the resulting meso- and macro-scale physics flows -- is the linchpin of any simulation campaign. Since no classical potential is universally transferable, one must choose from those available the potential that best captures the most pertinent physics. To faithfully model the pressure-induced $\alpha\to\omega$ phase transition in shock-loaded zirconium, we require the following of our potential: (1) that it correctly predict the high-pressure phase diagram of Zr; (2) that it reproduce the energy variation along the transition pathways between these phases; and (3) that it yield a reasonably accurate Hugoniot. The potential must also be computationally efficient enough to permit access to the `large' spatiotemporal scales of grain nucleation, growth, and coalescence. Taken together, these constraints leave perhaps only three Zr potentials \cite{Smirnova2017,Zong2018,Nitol2022} suitable for these high-pressure conditions.

From these viable potentials, we have selected the machine-learning-class potential developed by Zong \textit{et al.}\ \cite{Zong2018} (hereafter referred to as the Zong potential). In contrast to simpler, physically motivated potentials like those of the widely used embedded-atom-method (EAM) family, the Zong potential employs a complex and somewhat abstract parametrization scheme that affords it greater flexibility. In brief, it expresses the energy of each atom as a function of a set of unique configurational `fingerprints' distilled from its local bonding environment. The function that maps the local fingerprints onto a per-atom energy is constructed by training the potential on a multiphase energy database generated from \textit{ab initio} molecular dynamics (AIMD) simulations. This potential was tailored to reproduce the high-pressure allotropy of zirconium that we seek to model here, and has the added benefit of having already been tested in shock-loading scenarios \cite{Zong2019,Zong2020}, meaning we have previous results against which to benchmark our own.

As we show in Fig.~\ref{fig:phase_diagram}, the Zong potential largely succeeds in reproducing the experimentally measured \cite{Zhang2005} hydrostatic phase boundaries between the $\alpha$-, $\beta$-, and $\omega$-phases of Zr at moderate pressures. Its idealized Hugoniot in pressure-temperature space (the calculation of which is described in the Supplementary Material) also follows closely the locus of shock states predicted by the mutiphase equation of state (EOS) of Greeff \cite{Greeff2005}. The MD-simulated shock state we focus on here (which sits marginally above the idealized Hugoniot curve) lives deep in the $\omega$-stability region where the potential is expected to perform best.

However, the Zong potential has a defect that must be borne in mind throughout the following study: it is neither entirely smooth nor entirely continuous. This fact does not appear to be widely known; we are aware of only one other study documenting it, namely a very recent systematic comparison of thirteen Zr potentials by Nicholls \textit{et al.}~\cite{Nicholls2023}. The effect of these discontinuities is twofold: first, atoms sporadically experience abruptly amplified forces for periods of a few tens of femtoseconds at a time, thus modifying their dynamics (an example trajectory can be found in the Supplemental Material); second, truncation errors from the finite simulation timestep (1~fs) accumulate quickly enough that the atoms' total mechanical energy steadily increases over $\sim0.1$~ns timescales, meaning the evolution of the system is nonadiabatic. The latter problem can be mitigated (but not solved) by applying a gentle global thermostat that absorbs artificially generated heat, as we do here. However, this does not address underlying issue of unphysical atomic forces.

Nevertheless, we believe there \emph{is} value in examining the implications of the Zong potential, for two reasons. First, it is still possible to learn from physical models with known flaws. Indeed, we will show that, despite its shortcomings, the Zong potential correctly reproduces some salient features of experimental x-ray diffraction patterns from shock-compressed Zr. Second, it is important that we make known to the community the weaknesses of the Zong potential, and thus provide the motive to construct a more physically accurate and robust successor.

%  2.2 Simulation setup
\subsection{\label{subsec:setup} Simulation setup}

% Figure 2: Schematic of grow-and-prune scheme
\begin{figure}[b]
\centering
\includegraphics{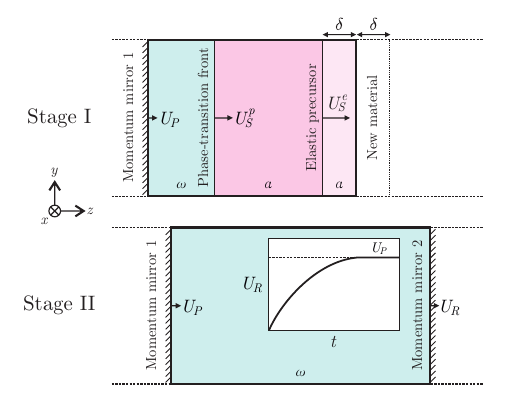}
\caption{\label{fig:grow-and-prune} Depiction of the grow-and-prune scheme used to conduct simulations of uniaxial shock-compression. In Stage I, a momentum mirror with speed $U_P$ drives a two-wave structure with an elastic precursor traveling at speed $U_S^e$. The crystal emerging from the mirror `grows' over the simulation; when the precursor wave comes within distance $\delta$ of the rear surface, the crystal is extended by the addition of fresh material of thickness $\delta$. At the start of Stage II, material beyond a specified cutoff is deleted (`pruned'). The remaining block of mostly compressed material nearest the first momentum mirror is isolated and its rear surface pressurized by a second momentum mirror traveling at speed $U_R\le U_P$. The inset illustrates the time-variation of $U_R$, which slowly approaches and eventually saturates at the particle velocity $U_P$.}
\end{figure}

Our molecular dynamics simulations largely follow the conventional template for modeling single crystals under uniaxial compression. The crystals we simulate are initially defect-free blocks of $\alpha$-Zr with their $[2\bar{1}\bar{1}0]$, $[01\bar{1}0]$, and $[0001]$ directions aligned with the $x$-, $y$-, and $z$-axes of the computational cell, respectively. The presence of laterally confining material surrounding the computational cell is imitated by the application of periodic boundary conditions (PBCs) in the $x$- and $y$-directions. Crystals span 40~nm in the transverse directions, which is more than sufficient to prevent nascent grains of the high-pressure $\omega$-phase interacting with periodic images of themselves at late times. Shock waves are launched by driving a momentum mirror into the lower $xy$-face of the crystal from rest to a final velocity of $U_p$ = 0.54~kms\textsuperscript{$-1$} (yielding a shock pressure of around 15~GPa) over a period of 0.8~ps (the rise time of the elastic precursor wavefront at this pressure). A simulation timestep of 1~fs is used throughout. All simulations are executed with the \textsc{lammps} code \cite{LAMMPS}; the Zong potential requires that we use the December 2014 release to ensure compatibility.

The Zong potential is around two orders of magnitude more expensive per atom-timestep than the workhorse EAM potentials commonly used in multimillion-atom shock simulations. Indeed, its cost is so great that the naive approach to shock-MD -- wherein a thick target is constructed with its final desired dimensions from the outset of the simulation -- is not only woefully inefficient, but also so expensive as to make an extended simulation campaign prohibitively unwieldy, even using supercomputing facilities. To make the campaign tractable, we need to reduce the simulation size without compromising the structure of the shock. Several modified integration schemes have been developed -- such as the Hugoniostat \cite{Ravelo2004} and multiscale shock technique (MSST) \cite{Reed2003} -- that impose uniaxial strain on a relatively small, homogeneous atomistic system in such a way as to respect the Rankine-Hugoniot jump conditions, thus eliminating the need for the large `runway' of material used in conventional shock-MD simulations. However, these schemes require the user to choose hyperparameters that influence the relaxation process and thus partially control the microstructure generated by the shock. Here, we use an alternative two-stage scheme similar in spirit to the moving-window technique \cite{Zhakhovsky2011} that allows us to efficiently simulate shock compression without the need to specify external parameters. We refer to our approach as the `grow-and-prune' scheme, which is illustrated in Fig.~\ref{fig:grow-and-prune}.

At the start of the simulation, there exists a sliver of $\alpha$-Zr spanning the full 40x40~nm\textsuperscript{2} transverse extent of the cell, but extending only 9~nm in the shock direction. During Stage I of the simulation, we `grow' the crystal along the compression direction by iteratively appending new material to its rear surface at a rate commensurate with the shock speed. A growth event is triggered when the foot of the leading shock front comes within distance $\delta = U_S^e\Delta t_{\text{eq}}$ of the crystal's rear surface, where $U_S^e$ is the speed of the elastic precursor wave and $\Delta t_{\text{eq}}\approx1$~ps is the time required for a pristine crystal to thermalize. At this instant, we append a new sliver of $\alpha$-Zr of thickness $\delta$ to the rear surface; the fresh material is able to equilibrate for time $\Delta t_{\text{eq}}$ before the shock front sweeps through it. By building the runway needed for the shock in this iterative manner, we avoid needless simulation of ambient material distant from the shock front and approximately double our computational efficiency.

% Figure 3: Lagrangian pressure profiles demonstrating grow-and-prune scheme
\begin{figure}[t]
\centering
\includegraphics{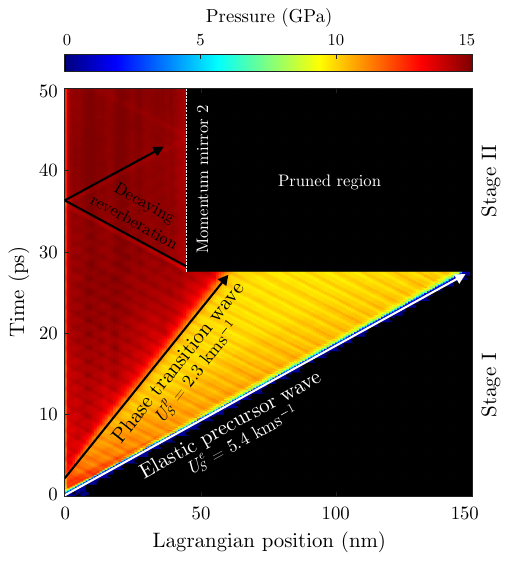}
\caption{\label{fig:pressure-surface} Spatiotemporal evolution of the pressure in single-crystal $\alpha$-Zr shock-compressed along [0001] with a momentum mirror travelling at 0.54 kms\textsuperscript{$-1$}, modeled using the Zong potential \cite{Zong2018}. New material is iteratively appended to the rear surface of the crystal to keep pace with the elastic precursor wave traveling at 5.4 kms\textsuperscript{$-1$}. At $t$ = 27.5~ps, material beyond $z_L$ = 45~nm is removed, and a second momentum mirror is installed at the crystal's new rear surface to keep the remaining material pressurized. Regions in which no material exists are rendered as black in this color scheme.}
\end{figure}

Once a sufficiently large and uniform body of compressed material has accumulated behind the shock, we select all atoms beyond this chosen region and remove them from the simulation; we refer to this as `pruning'. At the new free surface of the cell, we construct a second momentum mirror whose initial velocity $U_R(t_I)$ matches the local particle velocity of the rear surface. To disturb the wave structure as little as possible, we fit the time-variation of the particle velocity following the onset of the $\alpha\to\omega$ phase transition to a phenomenological sigmoid function, and then vary $U_R$ according to this function until it reaches the final particle velocity $U_P$, after which time we cap the second mirror's velocity:
\begin{equation}\label{eq:mirror-velocity}
    U_R (t) = \min\left[ U_0 + \Delta U \left(\frac{\frac{t-t_0}{\Delta t_{\text{front}}}}{1 + \left|\frac{t - t_0}{\Delta t_{\text{front}}}\right|}\right), U_P \right]\ .
\end{equation}
The fitting process and parameters are shown in the Supplementary Material. Once the second momentum mirror reaches speed $U_P$, the remaining material is effectively held under constant-volume conditions. The pruning step allows us to efficiently study slow, post-shock microstructural evolution at late times (hundreds of picoseconds after the shock) without needing to continue to simulate the shock wave itself.

To demonstrate the grow-and-prune technique, we show in Fig.~\ref{fig:pressure-surface} the structure of the pressure waves generated in our Zr shock-compression simulation. We observe a two-wave structure, with an elastic precursor wave traveling at $U_S^e$ = 5.4 kms\textsuperscript{$-1$} trailed by a phase-transition wave with velocity $U_S^p$ = 2.3 kms\textsuperscript{$-1$}. Note that, during Stage~I, successive crystal-growth events generate a train of weak stress waves that propagate back towards and eventually reflect from the front surface, leading to the conspicuous cross-hatch pattern visible in the region behind the phase-transition front. The magnitude of the pressure modulations amounts to no more than $\pm2.5\%$ of the local, time-averaged pressure, which we believe is too small to substantively change the dynamic response of the crystal. Indeed, we have verified that the behavior of Zr crystals simulated using a relatively inexpensive EAM potential \cite{Mendelev2007} with and without the iterative construction scheme are essentially identical (see Supplementary Material). We also note the presence of a reverberation of magnitude $\sim0.8$~GPa emitted at the beginning of Stage~II; it is difficult to eliminate the small velocity discontinuity introduced by the second momentum mirror entirely. However, the reverberation is unsupported and decays quickly, and is therefore unlikely to significantly change the microstructure of the final Hugoniot state.

The dynamic Stage~I of our simulations lasts 27.5~ps, while the quasistatic Stage~II lasts a further 172.5~ps. During the latter stage, we apply a Nos\'{e}-Hoover thermostat with a target temperature of 500~K (the average system temperature at the end of Stage~I) and a damping time of 5~ps, which is just sufficient to arrest the slow temperature drift caused by the artificially generated heat discussed in Sec.~\ref{subsec:potential}. The final shocked configuration we focus on here thus is found in the pressure-temperature state (15~GPa, 500~K).

We estimate (see Supplementary Material) that the grow-and-prune technique saves us a factor of 8 in computational costs. Without it, the cost of the Zong potential -- or other machine-learning-class potentials like it -- would render this simulation campaign unworkable.

%  2.3 Characterization techniques
\subsection{\label{subsec:characterization} Characterization techniques}

To characterize the real-space structure of the simulated atomistic configurations, we use a range of both standard and tailored coordination analysis techniques. The most important step in the workflow is the initial partitioning of the configuration into $\alpha$ (hcp), $\beta$ (bcc), $\omega$ (hex-3), and non-crystalline structures. In brief, the partitioning is realized in two steps. We first identify atoms in the $\omega$-phase with the Ackland-Jones parameter (AJP) \cite{AcklandJones2006} using a modified version of the original approach taken by Zong \textit{et al.}\ \cite{Zong2019}. We then classify the remaining atoms according to adaptive common-neighbor analysis (a-CNA) \cite{Stukowski2012b}. We give a detailed description of the full algorithm in the Supplemental Material. Note that we always perform the partitioning using atomic coordinates that have been time-averaged over 200~fs in order to combat thermal fluctuations, which degrade the accuracy of the structure classification.

To characterize the orientation of incipient crystal structures, we employ two different techniques. We measure the rotation (and elastic strain) state of the $\beta$-phase using the relatively computationally cheap template-matching technique (TMT) described in Ref.~\cite{Heighway2019}, whereby each atom's unit cell is assigned to one of a set of candidate unit cells based on the similarity of their orientations. For atoms in the $\omega$-phase (to which the TMT technique has not yet been extended), we instead use the per-atom structure factor (PASF) defined in Ref.~\cite{Higginbotham2013}, which is essentially the structure factor of a small spherical cluster of atoms surrounding the central atom. Originally used to discriminate between different twin variants in bcc tantalum, here we use the PASF to distinguish $\omega$ grains with different orientation relatonships with the original $\alpha$ crystal.

To synthesize diffraction patterns from the simulated atomistic configurations, we first obtain the raw ionic structure factor using the standard expression
\begin{equation}
S(\mathbf{q}) = \left|\sum_{\alpha=1}^N e^{-i\mathbf{q}\cdot\mathbf{r}_\alpha}\right|^2\ ,
\end{equation}
where $\mathbf{r}_\alpha$ is the instantaneous position of atom $\alpha$. We then sample the structure factor on the spherical locus of $q$-vectors constituting the Ewald sphere, and post-multiply the resultant intensity by the atomic-form, polarization, and self-attenuation factors. Note that whenever we calculate synthetic diffraction patterns from the late-time atomistic configuration, we Fourier transform not the entire cuboidal computational cell, but the inscribed sphere of atoms it contains. We do this to remove the hard, flat edges of grains at the edge of the box that would otherwise generate sharp, artificial, sinc-like features in reciprocal space.

% Figure 4: Visualisation of full computational cell
\begin{figure*}[t]
\centering
\includegraphics{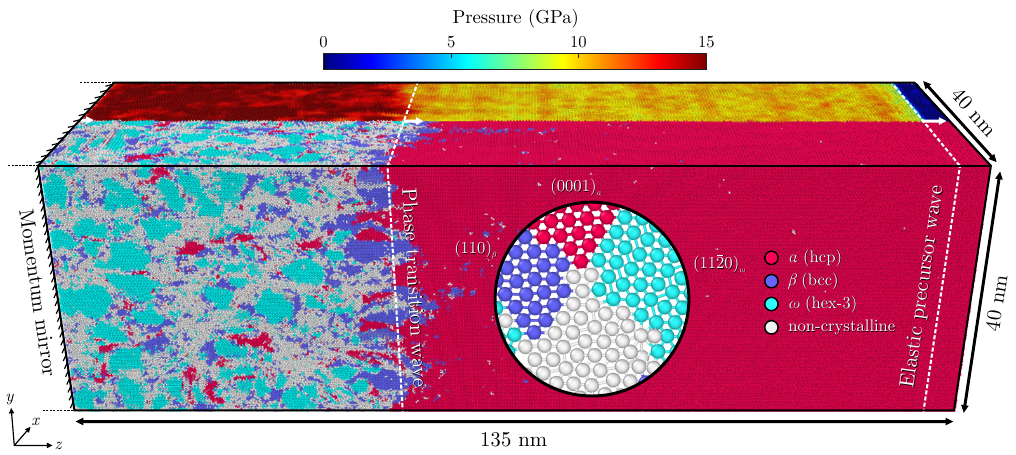}
\caption{\label{fig:cell_visualisation} Visualization of single-crystal Zr modeled under the Zong potential \cite{Zong2018} shock-compressed along its [0001] axis to 15~GPa using a momentum mirror with velocity $U_P = 0.54$ kms\textsuperscript{$-1$}. Foreground atoms are colored according to their local crystal structure, with $\alpha$, $\beta$, $\omega$, and non-crystalline environments rendered in magenta, blue, cyan, and gray, respectively. The inset shows a closeup of an intersection between the four structures viewed in a plane orthogonal to the shock. Atoms at the back of the cell are colored by their local pressure. This and all other visualizations were rendered in \textsc{ovito} \cite{Stukowski2010}.}
\end{figure*}

%  3. Results
\section{\label{sec:results} Results}

Our results are separated into real- and reciprocal-space analyses. The real-space analysis (Sec.~\ref{subsec:real-space}) focuses on the atomistic mechanisms by which the $\alpha\to\omega$ transition is realized and the complex microstructure that results. The reciprocal-space analysis (Sec.~\ref{subsec:reciprocal-space}) explores this microstructure's x-ray diffraction signatures -- both crystalline and diffuse -- and compares these features with those observed in femtosecond diffraction experiments on shock-compressed Zr.

%  3.1 Real-space analysis
\subsection{\label{subsec:real-space} Real-space analysis}

% Figure 5: Evolution of phase fractions
\begin{figure}[b]
\centering
\includegraphics{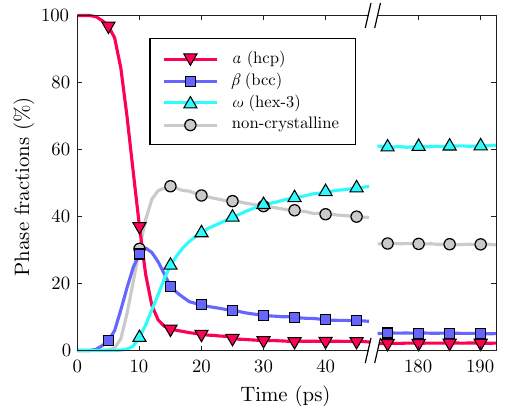}
\caption{\label{fig:phase_fractions} Evolution of the $\alpha$, $\beta$, $\omega$, and non-crystalline mass fractions in shock-compressed Zr following the arrival of the $\alpha\to\omega$ phase-transition front at $t=0$~ps. Calculations were performed on an 11-nm-thick Lagrangian material element containing one million atoms. Early- and late-time evolution are separated by a break in the time axis.}
\end{figure}

We begin by examining the gross structure of the shock wave. Shown in Fig.~\ref{fig:cell_visualisation} is an atomistic visualization of the shock-compressed Zr crystal after 27.5~ps, at the end of Stage~I of the simulation. At this instant, the computational cell is at its maximum extent (containing some 10 million atoms) and the two-wave structure has had time enough to fully develop. We color atoms by both their local pressure and crystal structure to correlate the wave structure with the underlying microstructural evolution.

An elastic precursor wave traveling at 5.4~kms\textsuperscript{$-1$} leads the shock, elevating the pressure to 10~GPa while preserving an $\alpha$ (hcp) crystal structure. The phase-transition front pursues the precursor wave at 2.3~kms\textsuperscript{$-1$}, and further raises the pressure to its final value of 15~GPa. Fig.~\ref{fig:cell_visualisation} reveals that the material behind the $\alpha\to\omega$ transition front possesses a rich, diverse microstructure comprising not only nanocrystals of the high-pressure $\omega$-phase, but also pockets of metastable $\beta$- (bcc-)Zr, in addition to swathes of interstitial non-crystalline matter. Our first objective is to identify the lattice-level mechanisms that create this complex configuration. It transpires that the short-lived $\beta$-phase -- an intermediate between the $\alpha$- and $\omega$-phases -- is absolutely essential to the formation of the rich microstructure we observe.

% Figure 6: Stages of grain evolution
\begin{figure}[t]
\centering
\includegraphics{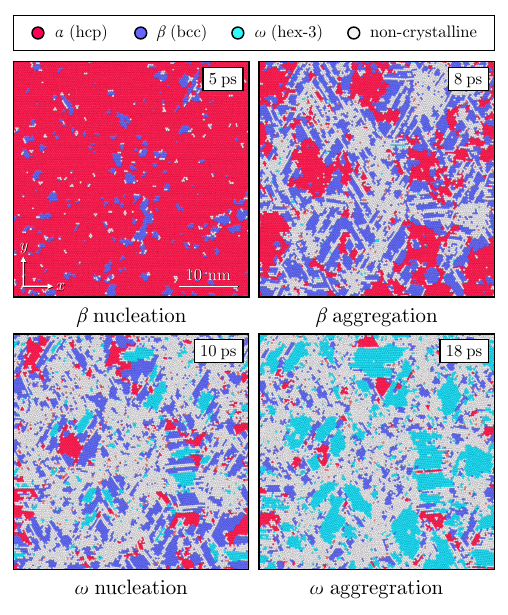}
\caption{\label{fig:cross-sections} Cross-sections of single-crystal Zr shocked along [0001] to 15~GPa, with atoms visualized by local crystal structure. Snapshots taken 5, 8, 10, and 18~ps after the arrival of the phase-transition front, marking the four stages of the $\alpha\to\beta\to\omega$ transition.}
\end{figure}

To better understand the post-shock phase dynamics, we show in Fig.~\ref{fig:phase_fractions} the evolution of the local mass fractions of the $\alpha$, $\beta$, $\omega$, and non-crystalline structures following the arrival of the phase-transition front. The parent $\alpha$-phase rapidly diminishes over a few picoseconds, and is initially replaced not by the $\omega$-phase, but the $\beta$. The nascent $\beta$-phase reaches its maximum mass fraction after just 10~ps, before decaying at the expense of a growing $\omega$-phase, whose population eventually saturates at just over 60\% of the crystal after 200~ps. The residual isolated clusters of untransformed $\alpha$- and $\beta$-Zr constitute only 2\% and 5\% of the late-time atomistic configuration, respectively. The remaining $33\%$ of the atoms assume non-crystalline structures. The majority of these atoms are situated on grain boundaries, but some form volumetric disordered nanoclusters sitting in the interstices between $\omega$ grains (as will be discussed later). The most important conclusion to be drawn from Fig.~\ref{fig:phase_fractions} is that there exists a brief but clearly defined `$\beta$ epoch' during which the bcc phase is the dominant extant crystal structure. We will now examine the $\alpha\to\beta\to\omega$ transition process in greater detail, and demonstrate the critical role the transient $\beta$ nanocrystal plays in the formation of the $\omega$ and non-crystalline material.

The transition process unfolds in four stages, which are illustrated in Fig.~\ref{fig:cross-sections}. First, supercritical $\beta$ nuclei form throughout the bulk of the uniaxially strained parent $\alpha$ crystal. The daughter nuclei undergo rapid, directional growth until they meet and form an extremely fine nanocrystalline aggregate. Only then do we observe the first appearance of $\omega$ nuclei. These (grand)daughter nuclei consume the $\beta$ grains as they grow until they meet and coalesce, thus forming a relatively coarse $\omega$ nanocrystal with a characteristic grain size of around 6~nm.

% Figure 7: Conventional UZ pathway
\begin{figure}[b]
\centering
\includegraphics{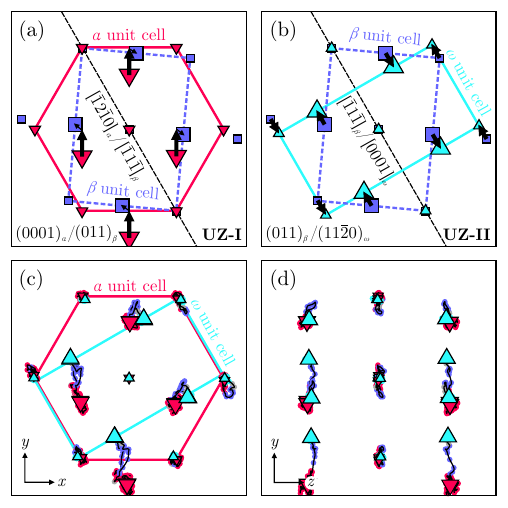}
\caption{\label{fig:uz-standard} Transformation of the Zr unit cell via the in-plane variant of the two-step Usikov-Zilberstein (UZ) pathway \cite{UZ1973}. (a) Atomic motion during the first stage (UZ-I), where the $\alpha\to\beta$ transformation is realized by shuffling of alternating $(0001)_\alpha$ planes in the $\pm[01\bar{1}0]_\alpha$ directions (indicated by black arrows). Atoms in different $(0001)_\alpha/(011)_\beta$ planes are distinguished by different marker sizes. (b) Atomic motion during the second stage (UZ-II), whereby the $\beta\to\omega$ transformation is accomplished by shuffling of successive $(\bar{2}\bar{1}1)_\beta$ planes along the $\pm[\bar{1}1\bar{1}]_\beta$ directions. (c) A typical trajectory in our shock MD-simulations, projected onto the $xy$-plane. Circular markers show atomic positions in the unit cell color-coded according to the instantaneous crystal structure of the cell; black traces have been added to make the trajectories more apparent. Larger markers indicate the time-averaged atomic positions during periods when the cell assumes $\alpha$ and $\omega$ structures. (d) The same trajectories projected onto the $yz$-plane.}
\end{figure}

% Figure 8: Omega nucleation sites
\begin{figure*}[t]
\centering
\includegraphics{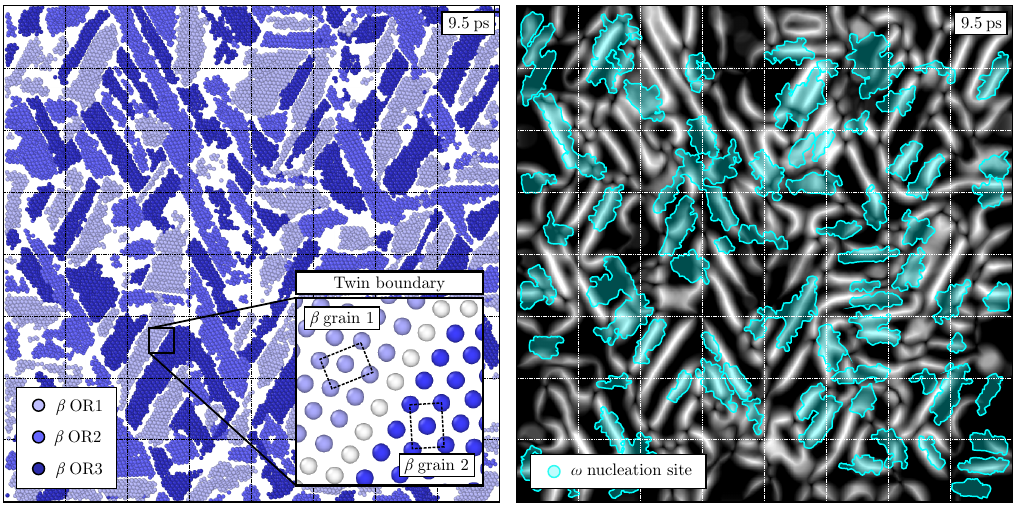}
\caption{\label{fig:omega_nucleation} Heterogeneous nucleation of the $\omega$ phase. (Left) $\beta$ atoms in a 2-nm-thick cross-section of the simulation cell, colored according to their approximate orientation relationship (OR) with the parent $\alpha$ crystal. Inset shows a coherent $\{112\}_\beta$ twin boundary between two $\beta$ grains of differing ORs. Image taken at $t=9.5$~ps, the height of the `$\beta$ epoch'.  (Right) Locations of $\omega$ nuclei formed between 9.0 and 13.0~ps overlaid on a map highlighting boundaries between $\beta$ grains at 9.5~ps.}
\end{figure*}

We have confirmed the prediction of this potential's authors \cite{Zong2019} that under the high-shear-stress conditions of uniaxial shock compression, the $\alpha\to\omega$ phase transition is realized almost exclusively by the `in-plane' variant of the Usikov-Zilberstein (UZ) mechanism \cite{UZ1973}. The two stages of this mechanism are illustrated in Figs.~\ref{fig:uz-standard}(a,b). The first stage (UZ-I) involves alternating $(0001)_\alpha$ planes shuffling along the $\pm[01\bar{1}0]_\alpha$ directions (along with modest compatibility shear and rotation) to convert $\alpha$-Zr to $\beta$-Zr via the Burgers mechanism. The second stage (UZ-II) sees the $\beta\to\omega$ transformation accomplished by shuffles along $\langle111\rangle_\beta$ directions orthogonal to the shock. In Figs.~\ref{fig:uz-standard}(c,d) we track the actual trajectories of the neighbors of an atom situated in a representative $\omega$ grain from our shock-MD simulation. The atomistic pathway they follow is identical, demonstrating that the dominant grain-growth mechanism is indeed the conventional in-plane UZ mechanism.

However, the mechanism by which the majority of the $\omega$ grains \emph{nucleate} is subtly different. To understand the nucleation process, we must first characterize the $\beta$ nanocrystal in which it unfolds. The symmetry of the Burgers mechanism by which the $\beta$-phase homogeneously nucleates is such that each $\beta$ grain can assume one of six orientations with respect to the parent $\alpha$ crystal \cite{Burgers1934}. The six orientations fall into three pairs of very similar orientations that we will not distinguish between, in order to simplify the discussion. The $\beta$ grains can thus be thought of as assuming three distinct orientations, all of which have their $(011)_\beta$ plane normal parallel to $(0001)_\alpha$ (i.e., the shock direction $z$), but which are related to one another by rotations of $120^\circ$ around $z$. We show on the left of Fig.~\ref{fig:omega_nucleation} a visualization of the $\beta$ grains at the instant when the $\beta$-phase is most populous, with grains colored according to which of the three orientation relationships (ORs) they take with respect to the $\alpha$ crystal. We observe that $\beta$ grains with dissimilar crystallographic orientations form a dense, interlocking structure with conspicuously straight boundaries that make angles of $0^\circ$, $120^\circ$, or $240^\circ$ with the $[2\bar{1}\bar{1}0]_\alpha$ direction (the $x$-axis).

Closer examination reveals that the grain boundaries are, in many cases, perfect $\{112\}_\beta$ twin boundaries (see inset of Fig.~\ref{fig:omega_nucleation}). This is a direct consequence of the symmetries of the Burgers mechanism, for which pairs of daughter $\beta$ orientations constitute exact reflections of one another in the $\{112\}_\beta$ planes orthogonal to $[011]_\beta$. When we superimpose the loci traced out by these twin boundaries and the locations of nascent $\omega$ grains formed during the first few picoseconds after the $\beta$ epoch (as we do on the right of Fig.~\ref{fig:omega_nucleation}) we see immediately that $\omega$ nucleation takes place preferentially on the twin boundaries between $\beta$ grains. That is to say that the $\omega$-phase nucleates \emph{heterogeneously}. After a supercritical $\omega$ nucleus forms on a $\beta$ twin boundary, it spreads into and consumes both $\beta$ grains either side of the boundary, generating a relatively large $\omega$ grain with a single orientation. Almost nowhere do we observe homogeneous nucleation of a $\omega$ crystallite within the bulk of a $\beta$ grain.

% Figure 9: Twinned UZ pathway
\begin{figure*}[t]
\centering
\includegraphics{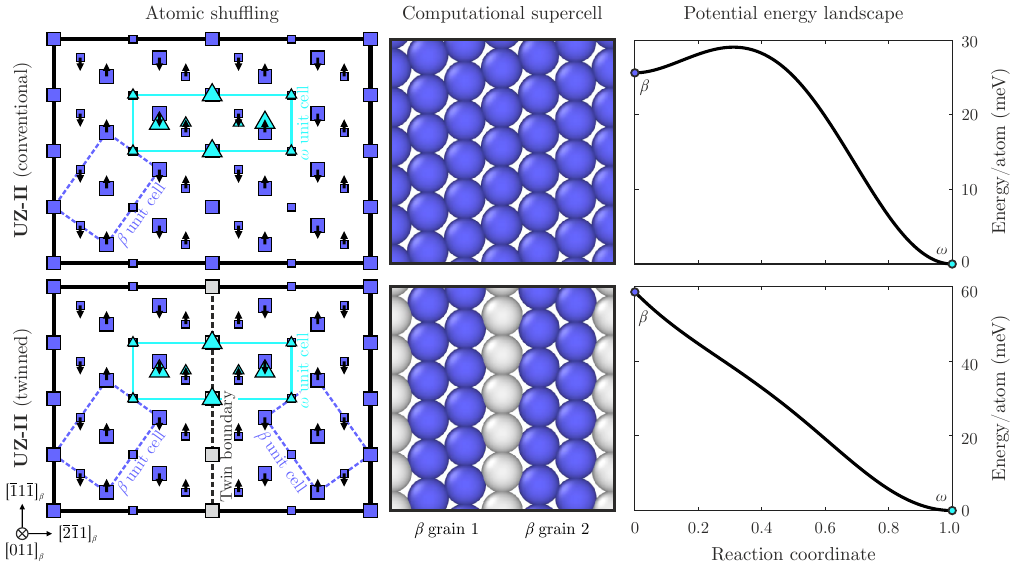}
\caption{\label{fig:uz-twinned} Comparison of the conventional (top) and twinned (bottom) variants of the second stage of the Usikov-Zilberstein pathway (UZ-II) mediating the $\beta\to\omega$ transition. Left panes illustrate the necessary shuffling of successive $(\bar{2}\bar{1}1)_\beta$ planes in the $\pm[\bar{1}1\bar{1}]_\beta$ directions with and without the presence of a coherent $(112)_\beta$ grain boundary. Middle panes show initial cross-sections of the prestressed supercell used to calculate the potential energy variation during UZ-II. Each fully periodic supercell contains 192 atoms, and is constructed with a cubic lattice constant of $a=3.5$~\AA\ and a compressive elastic strain of 7\% along $[011]_\beta$ to approximate the strain state of the metastable $\beta$ grains observed in shock-MD simulations. Right panels show the potential energy variation along each variant of the UZ-II pathway, defined with respect to the energy in the final $\omega$ state.}
\end{figure*}

To explain how and why heterogeneous $\omega$ nucleation occurs, we have considered the modified atomistic pathway by which the $\beta\to\omega$ transition would have to take place at a twin boundary. As illustrated in Fig.~\ref{fig:uz-twinned}, an $\omega$ crystal with the same orientation as that generated via the conventional UZ-II pathway can be formed on a coherent $(\bar{2}\bar{1}1)_\beta$ twin boundary by a simple reversal of direction of the $\pm[\bar{1}1\bar{1}]_\beta$ atomic shuffles at the $\{211\}_\beta$ twin boundary, i.e., at the same point where the $(\bar{2}\bar{1}1)_\beta$ stacking sequence reverses. We have directly compared the energy barriers acting against both the conventional and twin-modified variants of the UZ-II pathway using small-scale, static MD simulations of the supercells pictured in Fig.~\ref{fig:uz-twinned}. Each supercell contains a fully periodic $\beta$-Zr crystal with an initial elastic strain state akin to that of the transient $\beta$ nanocrystals observed in our shock-MD simulations. For the conventional UZ-II pathway, we observe that the metastable $\beta$ state is separated from the stable $\omega$ state by only a modest potential energy barrier (5~meV) easily overcome by thermal fluctuations, as anticipated. By contrast, the twinned UZ-II pathway exhibits no energy barrier whatsoever; the $\beta$ twin boundary is in fact \emph{unstable} to a transformation to the $\omega$ structure. While these simulations neglect the interface energy associated with the $\omega$ nucleus, and thus underestimate the kinetic barrier in both cases, we believe they nevertheless account for the observation that the coherent twin boundaries formed within the metastable $\beta$ nanocrystal act as preferential nucleation sites for the $\omega$ phase.

% Figure 10: Silcock vs TAO-I orientations
\begin{figure}[b]
\centering
\includegraphics{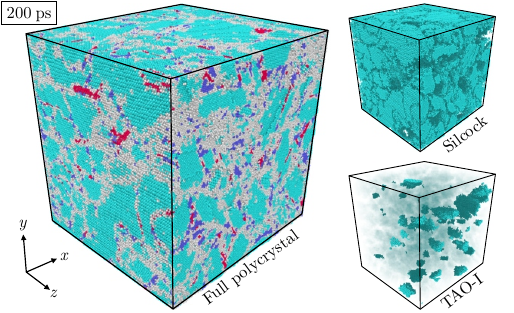}
\caption{\label{fig:silcock_vs_taoI} Late-time (200 ps) decomposition of the $\omega$ grains (cyan) into those with the Silcock and the TAO-I orientations (formed by the in-plane and out-of-plane variants of the Usikov-Zilberstein mechanism, respectively) using the per-atom structure factor \cite{Higginbotham2013}.}
\end{figure}

% Figure 11: Diffusion nucleation sites
\begin{figure*}[t]
\centering
\includegraphics{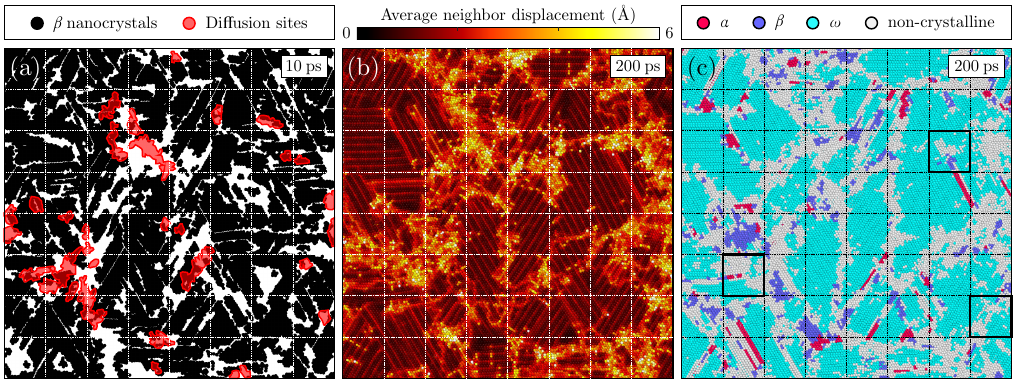}
\caption{\label{fig:diffusion-nucleation} Diffusive motion during the $\alpha\to\omega$ transition. (a) Locations of atoms at least one of whose neighbors move by more than 3.5~\AA\ within the first 4~ps of the $\beta$ nanocrystal aggregating, overlaid on a silhouette of $\beta$ grains at 10~ps. (b) Late-time configuration of the same set of atoms colored according to the average displacement suffered by their neighbors, highlighting atoms that participate in non-displacive motion. (c) The same atoms colored by their late-time local crystal structure. Boxed squares highlight $\omega$-structured material whose time-history is inconsistent with the displacive UZ pathway.}
\end{figure*}

We should note that while the in-plane UZ mechanism is dominant, we also observe a small proportion of $\omega$ grains that form via the `out-of-plane' variant, for which the UZ-II atomic shuffles take place in the $\langle111\rangle_\beta$ directions with a component parallel to the shock, rather than those orthogonal to it. The $\omega$ material with final orientations consistent with the in-plane mechanism (the Silcock orientation \cite{Silcock1958} for which $(0001)_\alpha\parallel(11\bar{2}0)_\omega$ and $[\bar{1}2\bar{1}0]_\alpha\parallel[0001]_\omega$) and with orientations consistent with the out-of-plane mechanism (the TAO-I orientation \cite{UZ1973} for which $(0001)_\alpha\parallel(01\bar{1}1)_\omega$ and $[11\bar{2}0]_\alpha\parallel[10\bar{1}1]_\omega$) constitute 95\% and 5\% of the late-time $\omega$ material, respectively. We show the morphology of the small and sparsely distributed TAO-I-oriented grains in Fig.~\ref{fig:silcock_vs_taoI}. For comparison, in the experimental study of Singh \textit{et al.}\ \cite{Singh2024}, the Silcock and TAO-I orientations respectively comprised 80\% and 10\% of the $\omega$ phase at 15~GPa, suggesting the Zong potential captures the transitions kinetics for the two corresponding transition pathways  reasonably well. However, we do not observe the third orientation detected by Singh \textit{et al.} (dubbed Variant-III), which comprised the remaining 10\% of the $\omega$-Zr. That the Zong potential fails to produce this third, previously unobserved pathway speaks to its lack of transferability (as commented upon in Ref.~\cite{Nicholls2023}).

Having studied the atomistic dynamics at the $\beta$ twin boundaries, it is natural to ask: what happens at the \emph{junctions} between $\beta$ grains? It is here that we see the fomentation of atomistic disorder. In Fig.~\ref{fig:diffusion-nucleation}(a), we highlight regions of a representative cross-section of the crystal in which atoms see at least one of their neighbors displaced by more than 3.5~\AA\ (a distance comparable to the typical atomic separation) within the first 4~ps of the $\beta$ nanocrystal reaching its maximum density. Such atoms cannot be participating in conventional displacive transition events like the UZ mechanism, but must instead be witnessing longer-range reordering of the kind seen in a diffusive transition. We observe that these diffusion sites preferentially appear at either incoherent, non-planar boundaries between $\beta$ grains, or at locations where several such grains converge, i.e., at locations where atoms are necessarily forced from the UZ pathway.

By the time the crystal approaches its late-time equilibrium state, the fraction of atoms that have participated in non-displacive motion has grown considerably. We show in Fig.~\ref{fig:diffusion-nucleation}(b) a map of the average displacement experienced by each atom's original nearest neighbors over the crystal's cross-section. Bodies of atoms that have undergone the regimented shuffling associated with the UZ mechanism are clearly marked out by their relatively small neighbor displacements, and by the equispaced striations that appear every third plane. Separating these bodies are veins of atoms whose average neighbor displacement is markedly greater. Approximately 35\% of all atoms have an average neighbor displacement greater than 1.5~\AA\ (approximately half the nearest-neighbor distance); around 10\% see at least one of their neighbors wander by distances of 5~\AA\ or more. The diffusive `channel' that operates alongside the conventional displacive one is thus highly active and forms an integral part of the microstructure formation process.

\begin{figure}[b]
\centering
\includegraphics{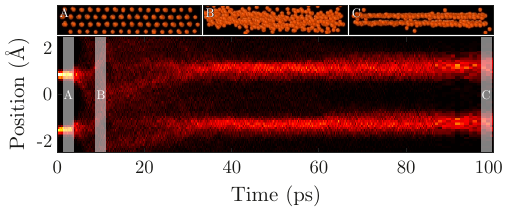}
\caption{\label{fig:diffusive-reconstitution} Linear number density of atoms undergoing diffusive restructuring in a $45\times45\times8$~\AA\textsuperscript{3} Lagrangian material element measured along the shock direction (the $z$-axis). Insets A, B, and C show side-on views (along the $x$-axis) of the element 2.5, 10, and 100~ps into the simulation, respectively, highlighting important stages of the $(0001)_\alpha\to(11\bar{2}0)_\omega$ restructuring.}
\end{figure}

When we compare the morphology of the displacive and diffusive bodies of atoms with that of the final crystal structure [shown in Fig.~\ref{fig:diffusion-nucleation}(c)], we observe a strong -- but not perfect -- correlation. The atoms whose environments transform in a displacive manner are those that assume the $\omega$ structure at late times (as expected) while nearly all atoms that suffer diffusive motion constitute the noncrystalline material at the boundaries or interstices between the $\omega$ grains. However, this rule does not hold universally: there exist pockets of material near the grain boundaries whose neighbor displacements are inconsistent with the simple displacive motion of the UZ mechanism, yet assume the $\omega$ structure nevertheless [see boxed regions in Fig.~\ref{fig:diffusion-nucleation}(c)]. To better understand the non-displacive dynamics in these boundary regions, we illustrate in Fig.~\ref{fig:diffusive-reconstitution} the time-history of the atomic plane structure in a representative material element. The atomic motion is clearly diffusive. We observe complete disintegration of the $(0001)_\alpha$ basal plane structure over a few picoseconds, followed by a gradual reconstruction of the $(11\bar{2}0)_\omega$ planes due to diffusive motion. Via this mechanism, atoms that stray from the UZ pathway can be `rescued' and allowed to join surface of an $\omega$ grain.

% Figure 12: Reformation of atomic planes via diffusion

Curiously, we find that the diffusive channel does not always yield a perfect $\omega$ structure. The volumetric pockets of non-crystalline material visible in Fig.~\ref{fig:diffusion-nucleation}(c) that live in the interstices between $\omega$ grains are the result of the diffusive channel forming an $\omega$-\emph{like} structure that is pervaded by self-interstitial defect atoms sitting between the $(11\bar{2}0)_\omega$ planes. Thus, the prevailing thermodynamic driving force towards the $\omega$-phase via the diffusive channel does not always succeed in creating an $\omega$ structure that is defect-free. This $\omega$-like material, which comprises about 5\% of the final atomistic configuration by mass, will be examined further in Sec.~\ref{subsec:reciprocal-space}.

To summarize, real-space analysis reveals that the shock-driven $\alpha\to\omega$ phase transition proceeds via a complex combination of both displacive and diffusive motion. The dominant process is the displacive Usikov-Zilberstein pathway, which involves the formation of an intermediate, short-lived $\beta$ structure via the Burgers mechanism. The majority of the granddaughter $\omega$ grains heterogeneously nucleate at twin boundaries in the transient $\beta$ nanocrystal via a modified version of the in-plane  UZ pathway, and assume the Silcock orientation with respect to the parent $\alpha$ crystal; the remainder transform via the out-of-plane UZ pathway and thus take the TAO-I orientation. Alongside conventional displacive reordering, we observe limited diffusive motion that is initiated at the junctions between $\beta$ grains and unfolds over tens to hundreds of picoseconds. This channel yields both defect-free and defective $\omega$ material, the latter being concentrated at the interstices between $\omega$ grains proper and constituting around 5\% of the configuration at late times. The $\omega$-phase comprises 60\% of the mass (of which 95\% and 5\% assume the Silcock and TAO-I orientations, respectively) and coexists with residual $\alpha$ and $\beta$ nanocrystals of mass fractions 2\% and 5\%, respectively. The remaining 28\% of the atoms form non-crystalline grain-boundary material.

%  3.2 Reciprocal-space analysis
\subsection{\label{subsec:reciprocal-space} Reciprocal-space analysis}

% Figure 13: Experimental vs simulated diffraction patterns
\begin{figure}[t]
\centering
\includegraphics{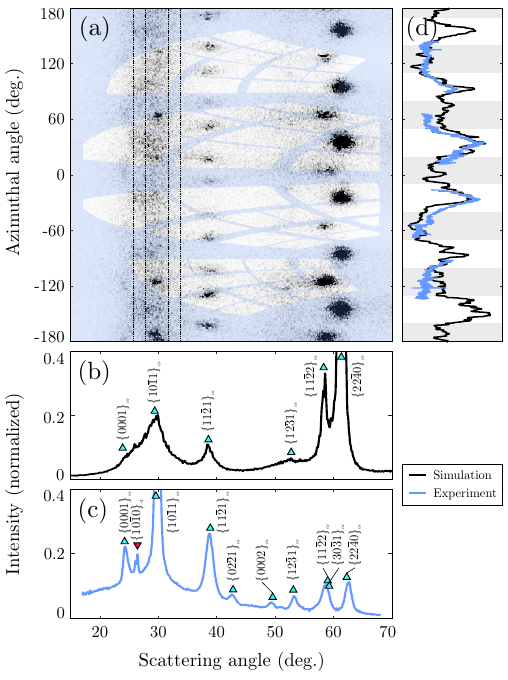}
\caption{\label{fig:diffraction-composite} Comparison of diffraction patterns from [0001]-oriented $\alpha$-Zr shocked to 15~GPa, generated from MD simulations (black) and obtained in a dynamic-compression experiment (blue)~\cite{Singh2024}, using a 10~keV x-ray beam at normal incidence to the sample. (a) Synthetic 2D diffraction pattern dewarped into $(2\theta,\phi)$-space, with scattering angle $2\theta$ varying horizontally and azimuthal angle $\phi$ vertically. The mask shows angular regions with no detector coverage in the experiment. (b,c) Azimuthally averaged diffraction patterns from the simulation and experiment, respectively. Intensities are expressed as a fraction of the maximum peak height. (d) Lineouts showing the average azimuthal variation of the diffuse scattering in the two thin bands centered around $2\theta=30^\circ$ in subfigure (a). Gaps in the experimental signal are caused by incomplete angular coverage of the detectors. Regions of maximal and minimal intensity in the diffuse band are indicated by white and gray bands, respectively.}
\end{figure}

While the real-space analysis in Sec.~\ref{subsec:real-space} gives us the luxury of examining the $\alpha\to\omega$ transition in atomistic detail, our understanding of this process is in reality limited to what can be discerned experimentally using ultrafast diagnostics, such as femtosecond x-ray diffraction (XRD). In this section, we will directly compare time-resolved diffraction patterns generated from our simulations with those obtained in dynamic-compression experiments, and interpret their form by examining the computational cell's full three-dimensional structure factor.

The experiment with which we will compare our simulations is described in full in the study of Singh \textit{et al}.~\cite{Singh2024}; we recap only the essential details here. The laser-driven shock-compression experiments were performed at the Matter in Extreme Conditions (MEC) \cite{MEC} endstation of the Linac Coherent Light Source (LCLS). The authors dynamically compressed single-crystal $\alpha$-Zr targets of 40~\textmu m thickness along their [0001] axis to pressures of between 5 and 18~GPa, using a 15-ns-long flat-top laser-pulse profile to ensure uniformity of the compressed state. Immediately before the shock front reached the rear surface of the targets, they were probed with a 50-fs burst of collimated 10~keV photons traveling parallel to the target normal (i.e., the original [0001] direction). The resulting time-resolved diffraction patterns were recorded on a set of four Cornell-Stanford Pixel Array Detectors (CSPADs) placed around the beam in a transmission geometry. Here, we focus on a shot taken on an $\alpha$-Zr target compressed to 15.2~GPa.

In Fig.~\ref{fig:diffraction-composite}(a), we show a synthetic two-dimensional diffraction pattern generated from our MD simulations at the latest time accessed (0.2~ns after compression begins) expressed in ($2\theta$,$\phi$) coordinates, where $2\theta$ is the scattering angle and where $\phi$ expresses the direction of scattering around the beam axis. The sixfold-symmetric diffraction pattern comprises a discrete set of scattering peaks distributed at regular intervals around a subset of the possible $\omega$-Zr Debye-Scherrer rings. The clearest rings are the $\{11\bar{2}1\}_\omega$ at $2\theta = 39^\circ$, the $\{11\bar{2}2\}_\omega$ at $59^\circ$, and the $\{22\bar{4}0\}_\omega$ (by far the brightest) at $61^\circ$. These peaks are primarily generated by the Silcock-oriented $\omega$ grains. There also exist several clear $\{10\bar{1}1\}_\omega$ Bragg peaks with irregular $\phi$-spacings at $2\theta = 30^\circ$, which come from the relatively rare TAO-I-oriented grains. Thus, of all possible reflections from an $\omega$-Zr polycrystal, we observe a highly restricted, discrete set due to the finite number of orientations the UZ pathway allows the $\omega$ grains to take.

When compared to the experimental pattern, the Bragg-like parts of the synthetic diffraction are similar in several aspects, and dissimilar in others. We show for reference the azimuthally integrated, one-dimensional patterns at 15~GPa in Figs.~\ref{fig:diffraction-composite}(b,c). For rings that are present in both patterns, we have confirmed that the peaks they contain appear at the same azimuthal positions. However, we note that a few reflections seen in the experiment -- most notably the extremely strong $\{10\bar{1}1\}_\omega$ peak at $30^\circ$ -- are largely absent in the simulated signal. These reflections are generated by a third set of crystallographic orientations of the $\omega$-phase observed by Singh \textit{et al.}\ that does not appear in our simulations. This, as discussed in Ref.~\cite{Singh2024}, suggests a lack of transferability of the Zong potential, and is perhaps unsurprising given that the atomistic pathway bringing about this third orientation relationship did not feature in this potential's training set.

The second major discrepancy is in the relative heights of the diffraction peaks. For example, while the $\{11\bar{2}2\}_\omega$ and $\{22\bar{4}0\}_\omega$ peaks are present in both the experimental and simulated patterns, they are far stronger in the latter. The discrepancy is owed to the peculiar nature of this particular experiment, in which a quasimonochromatic, collimated x-ray source probes a highly symmetric polycrystal containing only a small set of distinct crystal orientations. In such a scenario, there are, in general, no grains that meet the Bragg condition `exactly'; rather, all Bragg-like diffraction arises from near misses between the Ewald sphere and the maxima in the sample's structure factor. The intensity of such peaks is acutely sensitive to small changes in the crystal structure's density and orientation. It so happens for the present simulation that the $\{22\bar{4}0\}_\omega$ scattering vectors of grains with the Silcock orientation satisfy the Bragg condition almost perfectly (hence the extreme intensity of the $\{22\bar{4}0\}_\omega$ peak), while this is not the case in the experiment, for which the $c/a$ ratio is slightly different (0.605 compared with 0.621). This concept of marginal Bragg diffraction will be developed further shortly.

% Figure 14: Overview of reciprocal-space structure
\begin{figure*}[t]
\centering
\includegraphics{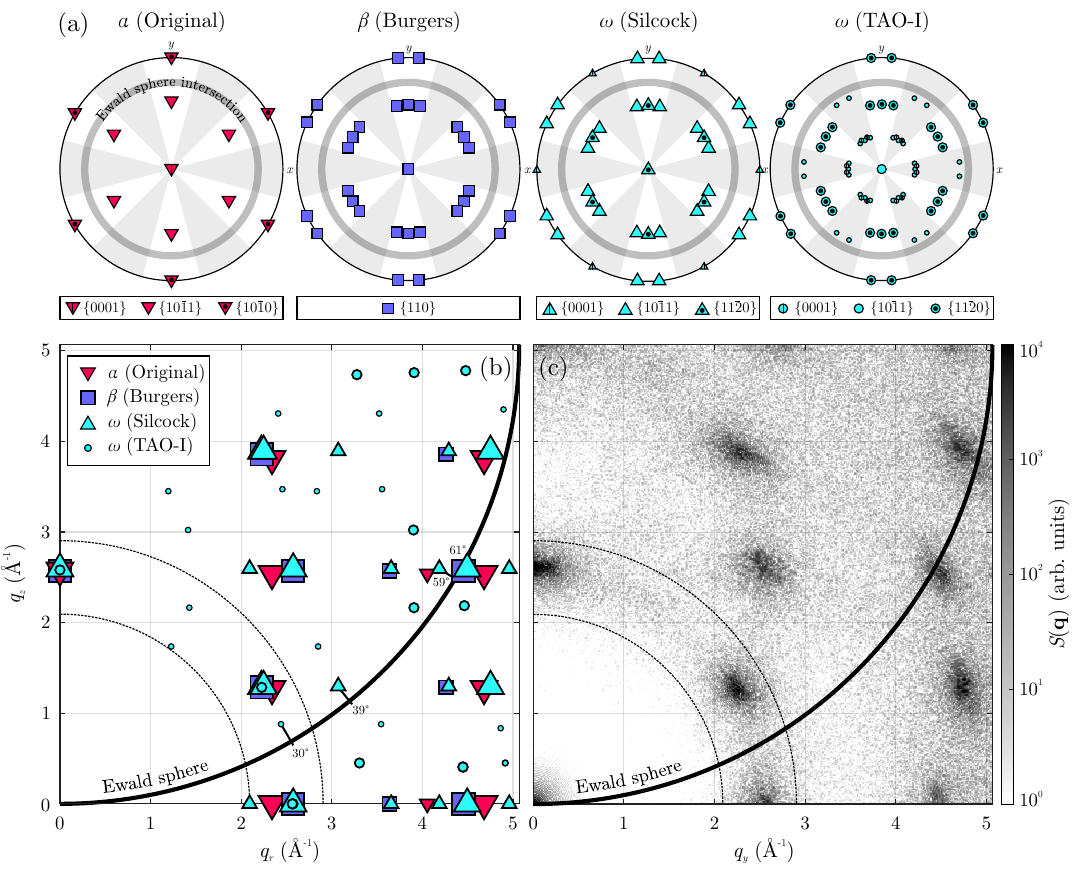}
\caption{\label{fig:reciprocal-space} Overview of the structure factor of the simulated $[0001]$-oriented $\alpha$-Zr crystal following the shock-induced $\alpha\to\omega$ phase transition at 15~GPa. (a) Pole figures showing ideal directions of scattering planes with $\mathbf{q}$-vectors satisfying $2.1\le\lVert\mathbf{q}\rVert\le2.9$~\AA\textsuperscript{$-1$}, including the original $[0001]$-oriented $\alpha$ single crystal, the 6 distinct $\beta$ Burgers orientations, the 3 distinct $\omega$ Silcock orientations, and the 12 distinct $\omega$ TAO-I orientations. Planes whose azimuthal direction falls near the maxima in the sixfold-symmetric diffuse scattering observed in the diffraction pattern (white sectors) are rendered with relatively large symbols; those near the minima (gray sectors) are smaller. Semitransparent rings show the regions intersected by the Ewald sphere (i.e., the regions visible in the diffraction pattern). (b) Collapsed view of reciprocal space showing reciprocal lattice vectors plotted with abscissa $q_r=\sqrt{q_x^2+q_y^2}$ and ordinate $q_z$ to demonstrate their proximity to the Ewald sphere. Symbols are again differentiated by size according to whether they fall in azimuthal regions containing maxima or minima of the diffuse scattering. Scattering vectors that actively contribute to Bragg-like peaks are connected to the Ewald sphere with their corresponding $2\theta$ angle labeled. Dashed lines indicate loci for which $\lVert\mathbf{q}\rVert = 2.1$~\AA\textsuperscript{$-1$} and 2.9~\AA\textsuperscript{$-1$}. (c) True structure factor of the target in the $q_yq_z$-plane (which contains a maximum in the azimuthal modulations). Intensity displayed on a logarithmic scale to expose interpeak structure.}
\end{figure*}

The feature of the diffraction on which we will focus in this study is not the Bragg-like scattering, but the anomalous diffuse background. In both the experimental and simulated one-dimensional diffraction patterns, we observe a broad band of scattering that is maximal at around $2\theta = 30^\circ$. A casual inspection of this signal might bring to mind either a liquid- or glass-like state, until one notices its azimuthal structure. As shown in Fig.~\ref{fig:diffraction-composite}(d), the diffuse band of scattering exhibits clear intensity modulations around the beam with a period of $\Delta\phi = 60^\circ$. The remarkable similarity between the structures of the diffuse band in the experiment and simulation indicates that the simulations may be bearing out the underlying physics correctly.

To better understand the structure of both the Bragg-like and diffuse components of the synthetic diffraction pattern, we have constructed an overview of the simulated specimen's full three-dimensional structure factor in Fig.~\ref{fig:reciprocal-space}. In Fig.~\ref{fig:reciprocal-space}(a), we provide pole figures showing the direction of planes in $\alpha$-, $\beta$-, and $\omega$-phases whose associated scattering vectors are of the appropriate magnitude to give scattering in the $2\theta\sim30^\circ$ region. We indicate in these pole figures the azimuthal regions containing the maxima in the diffuse scattering band by white sectors, and those nearer to minima by gray sectors, analogous to the shading in Fig.~\ref{fig:diffraction-composite}(d). Vectors at the center of the pole figures are directed along the shock direction, while those on the circumference are orthogonal to it.

We then show in Fig.~\ref{fig:reciprocal-space}(b) the reciprocal-space locations of these and many other scattering vectors collapsed onto a single plane containing the shock direction and the transverse direction. In this projection, reciprocal lattice vectors are plotted with coordinates
\begin{equation}
    (q_r,q_z) = \left(\sqrt{q_x^2 + q_y^2},q_z\right)\ ,
\end{equation}
such that all points on the same `orbit' around the shock axis appear at the same position. The purpose of this projection is to show the proximity of each reciprocal lattice vector to the Ewald sphere and hence its degree of contribution to the diffraction pattern, for which one only needs to know its cylindrical coordinates $q_r$ and $q_z$. We differentiate between scattering vectors with different azimuthal positions around the shock direction only by symbol size: those situated in regions nearer the maxima in the diffuse band [i.e., those in white sectors in Fig.~\ref{fig:reciprocal-space}(a)] are plotted with larger symbols for emphasis, while those nearer the minima [in gray sectors in Fig.~\ref{fig:reciprocal-space}(b)] have smaller symbols. For clarity, scattering vectors that contribute appreciably to the diffraction pattern due to their proximity to the Ewald sphere are labeled by their corresponding $2\theta$ angle. Together, Figs.~\ref{fig:reciprocal-space}(a,b) explain the Bragg-like structure of the diffraction pattern in Fig.~\ref{fig:diffraction-composite}.

We first reiterate the observation that the Bragg condition is met perfectly almost nowhere. The $\{10\bar{1}1\}_\omega$ peak at $30^\circ$ from the TAO-I-oriented grains and the $\{11\bar{2}1\}_\omega$ and $\{11\bar{2}2\}_\omega$ peaks at $39^\circ$ and $58^\circ$ from the Silcock-oriented grains are all generated by scattering vectors sitting marginally `below' the surface of the Ewald sphere. It is only the $\{22\bar{4}0\}_\omega$ scattering vectors -- and, in fact, the $\{220\}_\beta$ vectors, with which they overlap -- that satisfy the Bragg condition almost perfectly, leading to its exceptionally strong diffraction peak.

We next note that, with the exception of the $\{10\bar{1}1\}_\omega$ scattering vectors from the TAO-I-oriented grains, the region of reciprocal space in which the diffuse scattering is maximal (namely the $2\theta\sim30^\circ$ zone of the Ewald sphere) is devoid of coherent scattering features. In this region, the Ewald sphere threads its way through a gap flanked by clusters of nearly overlapping reciprocal lattice vectors from the $\alpha$-, $\beta$-, and $\omega$-phases. The cluster outside the Ewald sphere comprises a $\{10\bar{1}0\}_\alpha$ peak along with pairs of $\{110\}_\beta$, $\{10\bar{1}1\}_\omega$, and $\{11\bar{2}0\}_\omega$ vectors, all of which sit at $90^\circ$ to the shock direction. The cluster inside the Ewald sphere comprises a $\{10\bar{1}1\}_\alpha$ peak along with triplets of almost exactly overlapping $\{110\}_\beta$, $\{10\bar{1}1\}_\omega$ and $\{11\bar{2}0\}_\omega$ peaks inclined at around $60^\circ$ to the shock direction. As illustrated in the pole figures in Fig.~\ref{fig:reciprocal-space}(a), the scattering vectors in question fall exclusively in azimuthal regions coinciding with the maxima in the diffuse scattering band. This hints at the possibility that the diffuse signal is the result of the Ewald sphere sampling the tenuous wings of the intensity distribution around these clusters of scattering vectors.

To show this directly, we have plotted in Fig.~\ref{fig:reciprocal-space}(c) the structure factor of the atomistic configuration calculated from its Fourier transform. The structure factor is shown in the $q_yq_z$-plane, meaning the visible maxima correspond to the reciprocal lattice vectors plotted with larger symbols in Fig.~\ref{fig:reciprocal-space}(b). We observe that the two previously mentioned clusters of scattering vectors centered approximately at coordinates (2.55,\,0)~\AA\textsuperscript{$-1$} and (2.25,\,1.25)~\AA\textsuperscript{$-1$} are connected by a bridge of non-negligible scattering intensity through which the Ewald sphere passes. The value of the structure factor in this diffuse region is 2 to 3 orders of magnitude smaller than at the maxima. Diffuse scattering of this magnitude would ordinarily be eclipsed by coherent scattering from nearby Bragg peaks, were it not for the sparse nature of the target's reciprocal space. What, then, is the origin of the tenuous interpeak scattering intensity?

% Figure 15: Effect of quenching diffraction pattern
\begin{figure}[t]
\centering
\includegraphics{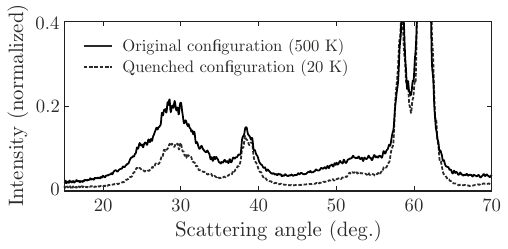}
\caption{\label{fig:diffraction-quench} Synthetic diffraction patterns from [0001]-oriented $\alpha$-Zr shocked to 15~GPa before and after quenching from the original 500~K shock state to one with a temperature of just 20~K. Both patterns are normalized by the maximum peak height from the 500~K signal. The weak $\{10\bar{1}\}_\omega$ Bragg peak at $30^\circ$ have been masked out to make clearer the shape of the diffuse scattering peak sitting below it.}
\end{figure}

Several mechanisms can siphon intensity from the crystalline maxima of the structure factor into the space between them. Candidates include: (1) the displacement of atoms from their ideal crystal sites due to thermal motion, which causes thermal diffuse scattering (TDS); (2) the presence of extremely small coherently scattering domain sizes, whose reciprocal-space extent is relatively large; and (3) disordering due to high densities of crystal defects. It transpires that all three of these mechanisms contribute to the simulated diffuse scattering signal.

To measure the amount of TDS in our simulations, we compared the structure factor of the original atomistic configuration with that of another configuration obtained by quenching the first, i.e., one whose temperature has been rapidly reduced by a thermostat. We compare the synthetic diffraction of the `hot' (500~K) and `cold' (20~K) configurations in Fig.~\ref{fig:diffraction-quench}. Here, we have masked out the $\{10\bar{1}1\}_\omega$ peaks from the TAO-I-oriented grains that sit atop the diffuse band to make the latter's shape and magnitude clearer. We observe that while quenching the system leaves the height of the Bragg peaks largely unchanged (in fact, the $\{22\bar{4}0\}_\omega$ peak noticeably \emph{grows} thanks to the reduced Debye-Waller effect), the diffuse signal falls to approximately half of its original value over the entire sampled $2\theta$ range. It should be acknowledged that the phonons in our classical simulations are necessarily distributed according to Maxwell-Boltzmann rather than a Bose-Einstein statistics, and therefore the absolute values of the TDS that our simulations predict should be treated with caution. With that said, we note that the simulated TDS accounts for only half of the diffuse signal; even after quenching, there remains a diffuse peak whose integrated intensity is still comparable to that of the Bragg peaks. The remaining non-thermal intensity must be explained either by very small scattering domains (in the form of the residual $\alpha$ and $\beta$ nanograins), or the presence of material lacking crystalline order.

Cluster analysis of the residual $\alpha$ and $\beta$ phases (which have mass fractions of 2\% and 5\%, respectively) reveals that their typical grain diameter is of order $D$ = 2-3~nm. The dimension of the shape function dressing each of their scattering vectors is therefore $2\pi/D \sim$ 0.2-0.3~\AA\textsuperscript{$-1$}, a distance comparable to the separation between the above-mentioned clusters of scattering vectors and their points of closest approach on the Ewald sphere. In other words, the Ewald sphere must be sampling non-negligible intensity from $\alpha$ and $\beta$ nanoparticles' broad structure factor, thus contributing to the (anisotropic) diffuse scattering signal.

\begin{figure}[t]
\centering
\includegraphics{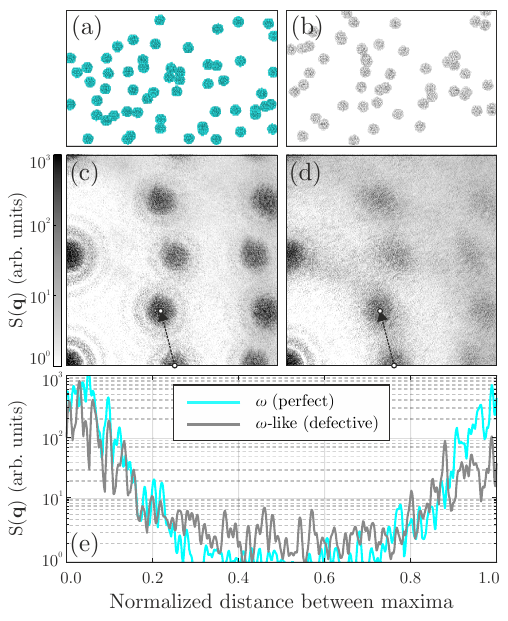}
\caption{\label{fig:clusters} Comparison of the structure factors of perfect $\omega$ and defective $\omega$-like nanoclusters taken from the late-time atomistic configuration. (a,b) Visualization of the $\omega$ and $\omega$-like nanoclusters, respectively, all with a radius of approximately 2~nm. (c,d) Corresponding structure factors calculated in the $q_yq_z$-plane, displayed on a logarithmic scale. (e) Comparison of the structure factor variation along the path between the maxima indicated by dashed arrows in subfigures (c,d).}
\end{figure}

Similarly, we identified in Sec.~\ref{subsec:real-space} pockets of long-lived, defective, $\omega$-like material occupying the interstices between $\omega$ grains. That these atoms are relatively poorly ordered suggests their structure factor may be of inherently `lower contrast', with appreciable intensity between the Bragg peaks. To compare the structure of this $\omega$-like material with that of the perfect $\omega$ crystal in a like-for-like manner, we carved out around 50 spherical nanoclusters of each structure from the unquenched atomistic configuration [see Fig.~\ref{fig:clusters}(a,b)], and calculated their structure factors normalized by the total number of atoms sampled. Crucially, the nanoclusters all have the same diameter of 2~nm (clusters of $\omega$-like material larger than this are rare). The identical cluster size means the extrinsic broadening of the structure factor peaks caused by the finite scattering domain size is equivalent, and any remaining broadening must be intrinsic to the atomistic structure.

% Figure 16: Comparison of omega and omega-like structure factors

We compare the structure factors of the $\omega$ and non-crystalline $\omega$-like nanoclusters sampled in the $q_yq_z$-plane in Figs.~\ref{fig:clusters}(c,d). The structure factor of the perfect $\omega$ clusters has the expected form, namely a discrete set of points dressed by the oscillatory, spherically symmetric shape functions characteristic of a small, spherical scattering domain. The structure of the defective $\omega$ material is qualitatively similar, but is of much lower contrast. The intensity of the peaks diminishes considerably faster with increasing $q$, and far more intensity is distributed between the peaks. To make this more evident, we show in Fig.~\ref{fig:clusters}(e) the intensity variation along the reciprocal-space path shown in Figs.~\ref{fig:clusters}(c,d), which traverses the region whence the diffuse signal comes. While the difference between the coherent and diffuse scattering is a factor of almost 1000 for the $\omega$ nanoclusters, it is only 10-100 for the defective material. We speculate that the diffuse scattering seen in the synthetic diffraction patterns arises partially from the tenuous wings of the maxima in the structure factor of the interstitial, partially disordered material.

The question that immediately arises is: what is the relative contribution of the $\alpha$ and $\beta$ nanoparticles and the interstitial defective nanoclusters to the (non-thermal) diffuse scattering? The natural way to evaluate the contribution of (say) the defective material would seem to be to calculate synthetic diffraction from the disordered nanoclusters in isolation, with all other atoms removed, and to calculate the fraction of diffuse peak it accounts for. In practice, we find that the diffuse feature produced by these defective pockets of material alone is of considerably greater width than the diffuse band from the complete aggregate; the same is true when we isolate the $\alpha$ and $\beta$ nanoparticles. Indeed, we have made many attempts to additively decompose the diffraction pattern into separate contributions from the $\omega$ grains, the partially disordered interstitial material, and the residual $\alpha$ and $\beta$ grains, and we have invariably failed. The reason is that we cannot assume the positions of atoms in adjacent regions of differing phases are uncorrelated. Looking again at Fig.~\ref{fig:reciprocal-space}, we see that there are many instances of reciprocal lattice vectors from the $\alpha$-, $\beta$-, and $\omega$-phases overlapping almost exactly. This follows from the shuffle-mediated nature of the $\alpha\to\beta\to\omega$ phase transition, which preserves the crystal's planar structure in certain directions despite causing drastic changes in atomic coordination. The clearest instantiation of this effect is the spike of reciprocal-space intensity concentrated at $\mathbf{q}=(0,0,2.6)$~\AA\textsuperscript{$-1$}, which represents a near-perfect overlap of the $(0001)_\alpha$, $(011)_\beta$, $(11\bar{2}0)_\omega$ (Silcock), and $(10\bar{1}1)_\omega$ (TAO-I) scattering vectors; to ignore coherence and treat the aggregate peak as a linear superposition of peaks from individual phases would be incorrect. In a similar vein, correlations between the positions of atoms in the $\omega$ grains and those of the partially disordered $\omega$-like pockets between them cannot be ignored. These phases (if the $\omega$-like material can be described as such) are contiguous, and have structures differing only by the presence or absence of point defects. It therefore does not make sense to treat the two subsets of atoms as independent scattering bodies; one can only speak of scattering from their aggregation.

For these reasons, we cannot partition the non-thermal component of the diffuse scattering into contributions from defective, partially disordered material and from the nanoscale $\alpha$ and $\beta$ grains that survive the transition process, and we further contend that doing so is impossible; while both effects surely contribute, they do so nonlinearly. It is only with an atomistic simulation technique such as molecular dynamics that the minutiae of the correlations between atomic positions in the various daughter phases -- and their effect on the form of the diffuse scattering signal -- can be faithfully modeled.

The final point we shall make is that it is largely by virtue of the specimen's sparse reciprocal space that the anisotropic diffuse scattering is so conspicuous. We previously suggested that the relatively weak diffuse signal would be obscured by conventional Bragg-like scattering for the kind of polycrystalline samples more regularly used in femtosecond x-ray diffraction experiments. To generate a representative diffraction pattern from a powderlike sample without simulating a true polycrystal under dynamic compression (which would be exorbitantly expensive under the present potential), we take the full three-dimensional structure factor from our single-crystal simulations, $S(\mathbf{q})$, and calculate its one-dimensional polycrystalline analogue using an unweighted spherical average:
\begin{equation}
    \langle S \rangle (q) = \frac{1}{4\pi} \int d\Omega\, S(\mathbf{q})\ .
\end{equation}
The resulting structure factor is that of an ensemble of independently scattering replicas of our final configuration with an isotropic orientation distribution. By taking this average, we implicitly assume that every parent $\alpha$ crystallite, no matter its initial orientation with respect to the compression direction, reaches a shock state with an identical (albeit rotated) microstructure, which would almost certainly not be the case in reality. We do not believe this detail is important for the purposes of the comparison we wish to make, as it is only the leading-order relative intensities of the Bragg-like and diffuse scattering that matter.

% Figure 17: Comparison of mono- and polycrystalline diffraction
\begin{figure}[t]
\centering
\includegraphics{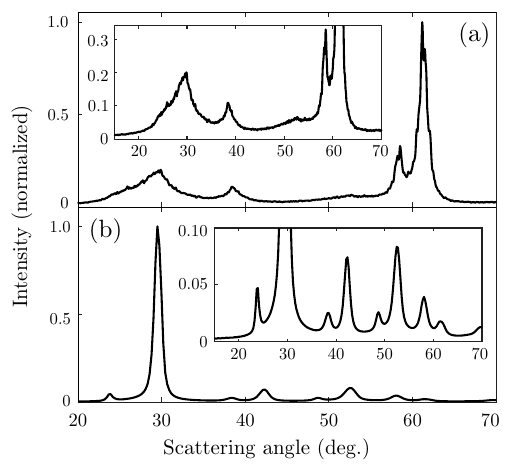}
\caption{\label{fig:powder} Comparison of synthetic diffraction patterns of (a) the original, intially monocrystalline atomistic configuration and (b) a powderlike polycrystalline analogue obtained by spherical averaging of the former configuration's structure factor. Insets show closer views of the lower-intensity peaks for each configuration. Intensities are expressed as a fraction of the maximum peak height.}
\end{figure}

Figure \ref{fig:powder} compares the azimuthally integrated diffraction patterns from the original single-crystal shock-MD simulation with that of a polycrystalline analogue. The difference is stark. In the former case, the $\{22\bar{4}0\}_\omega$ at $2\theta=61^\circ$ dominates the landscape, and the diffuse peak centered at $30^\circ$, while of far lower maximal intensity, is clearly resolvable and of a comparable integrated intensity. For the polycrystalline analogue, meanwhile, all other reflections are dwarfed by an exceptionally intense Bragg peak at $29^\circ$, which is in fact a combination of closely overlapping $\{10\bar{1}1\}_\omega$ and $\{11\bar{2}0\}_\omega$ Debye-Scherrer rings. Resolving a weak diffuse peak beneath the towering $\{10\bar{1}1\}_\omega/\{11\bar{2}0\}_\omega$ Bragg peak would be extremely difficult without strong prior constraints on the latter's lineshape. We therefore maintain that the anisotropic diffuse scattering signal is visible thanks to the single-crystal nature of the target. This concept will be discussed further in Sec.~\ref{sec:discussion}.

In summary, we have generated synthetic diffraction patterns from [0001]-oriented single-crystal $\alpha$-Zr shock-compressed to 15~GPa following the $\alpha\to\omega$ phase transition using a 10~keV x-ray beam at normal incidence to the simulated target. In addition to a subset of the full complement of $\omega$ diffraction peaks, we observe an anisotropic diffuse scattering signal centered at $2\theta = 30^\circ$ with sixfold rotational symmetry around incident beam direction, much like that observed by Singh \textit{et al.}\ in an analogous dynamic-compression experiment \cite{Singh2024}. The (classical) simulations suggest that approximately half of the diffuse signal arises from thermal diffuse scattering, while the remainder is generated by a combination of interstitial, defective $\omega$-like material and residual, nanoscale $\alpha$ and $\beta$ grains that survive the transition. We show that the diffuse peak would likely be obscured by conventional Bragg-like scattering were the target's initial texture to be polycrystalline.

%  4. Discussion
\section{\label{sec:discussion} Discussion}

While the present computational study makes a number of interesting predictions about the physics of zirconium under dynamic compression, we believe its most important predictions pertain to the peculiarities of the single-crystal diffraction configuration. The [0001]-oriented $\alpha$-Zr single crystals used in the experimental study of Singh \textit{et al.}~\cite{Singh2024} that we simulate here possess the strongest starting texture possible. In other words, their reciprocal space is populated by a sparse set of extremely localized scattering vectors, and is otherwise entirely empty. Such targets are therefore virtually guaranteed not to satisfy the Bragg condition when illuminated by a collimated, quasimonochromatic x-ray source. We have shown that even following the Usikov-Zilberstein-mediated $\alpha\to\omega$ phase transition, which creates $\beta$- and $\omega$-Zr grains with 6 and 15 distinct orientations, respectively, the specimen's reciprocal space remains sparse in real terms. By this, we mean that the Bragg condition is exactly satisfied virtually nowhere, except when the combination of shock-induced strains and rotations happens to map a scattering vector onto the Ewald sphere. We observed this behavior in the $\{22\bar{4}0\}_\omega$ peak at 15~GPa, but changing the shock pressure even slightly could displace this scattering vector from the Ewald sphere and thus largely extinguish its Bragg peak. The majority of the diffraction peaks are generated by scattering vectors that miss the Ewald sphere by varying degrees.

The consequence of the target's sparse reciprocal space is that the overall intensity scale of the diffraction is vastly reduced, and the off-Bragg, diffuse elements of the diffraction pattern that might ordinarily be overlooked (or even subtracted off as a background) suddenly come to the fore. By way of comparison, we approximated the diffraction pattern generated by a powderlike polycrystalline Zr target, and demonstrated that the diffuse peak at $30^\circ$ would likely be buried beneath the wings of the `true' Bragg peaks. This appears to account for the fact that the diffuse features observed in our simulations are also observed in the single-crystal study of Singh \textit{et al.}, but \emph{not} in analogous experiments performed using polycrystals \cite{Singh2024}. To reiterate, it not necessarily the case that the diffuse scattering in this configuration is particularly `strong'; rather, the Bragg scattering is `weak'. Note that this phenomenon is by no means peculiar to zirconium; one could reasonably expect that diffuse scattering of a qualitatively similar nature might be observable in \emph{any} single crystal whose texture is substantively diversified by dynamic compression.

With that being said, there are aspects of the present experiment that make the diffuse signal more conspicuous. First, the fact that the diagnostic x-ray beam travels at normal incidence to the target normal -- which is to say, parallel to the target's initial axis of sixfold symmetry -- is not unimportant. The orientation (and radius) of the Ewald sphere is such that its $2\theta\sim30^\circ$ zone exclusively samples regions of particularly strong diffuse scattering that appear every $\Delta\phi=60^\circ$ around its surface. These regions are largely devoid of coherent scattering features (with the exception of the $\{10\bar{1}1\}_\omega$ from the TAO-I-oriented grains) and moreover correspond to relatively modest Bragg and emergence angles, such that attenuation due to the atomic-form and self-attenuation factors in this zone is weak. These effects conspire to give a clearly defined diffuse signal that becomes clearer yet after azimuthal averaging. Bringing the x-rays in at a non-zero incidence angle would break the diffraction pattern's sixfold rotational symmetry and thus blur the structure of the diffuse scattering. Tilting the Ewald sphere would also increase the `probability' of one or more of the multitude of scattering vectors satisfying $\lVert\mathbf{q}\rVert\sim2.5$~\AA\textsuperscript{$-1$} meeting the Bragg condition and thus overshadowing the diffuse signal.

Second, the molecular dynamics simulations suggest that $\alpha$-Zr dynamically compressed along [0001] contains microstructures that generate anisotropic diffuse scattering over and above conventional thermal diffuse scattering. Under the Zong potential, the $\alpha\to\omega$ phase transition is realized primarily via the displacive Usikov-Zilberstein mechanism, which, as described in Sec.~\ref{subsec:real-space}, brings about the formation of a short-lived, extremely fine $\beta$-Zr nanocrystal. The atomistic disorder fomented at the incoherent boundaries between the $\beta$ nanograins ultimately leads to the diffusive formation of pockets of defective $\omega$-like material whose structure factor is characterized by non-negligible intensity between the peaks. The simulations further suggest that a subset of the $\beta$ grains is kinetically inhibited from completing the transformation to the $\omega$-phase (over the 0.2~ns simulation timescale, at least), and that these minuscule grains act as an additional source of relatively delocalized diffraction. Within the classical framework of the simulations, the thermal and non-thermal parts of the diffuse scattering are of almost equal strength.

An essential question is whether the diffuse signal can be used to diagnose the dynamically compressed target's Hugoniot state: the thermal component is directly linked to the system's temperature, the non-thermal component to the mesoscale structure and defect content. Despite our best efforts, we have found that it is not possible to additively decompose the non-thermal diffuse scattering into contributions from the residual $\alpha$ and $\beta$ grains and from the defective $\omega$-like material. Furthermore, although the thermal diffuse scattering \emph{is} apparently separable, its structure is very similar to that of the non-thermal diffuse scattering. To `invert' the diffuse signal and decompose it into its constituent parts is, therefore, not a trivial task. Molecular dynamics simulations do at least allow us to model the diffuse scattering in a forward sense and to check for consistency with experimental data. Indeed, we believe atomistic simulation techniques are \emph{required} to faithfully model diffuse scattering of the kind observed here, as they automatically incorporate the subtleties of atom-atom correlations within and between different phases in a way that meso- or continuum-scale models usually do not.

Of course, the fidelity of these simulations depends on the interatomic potential they use. We have shown that the Zong potential \cite{Zong2018} bears out several salient features of the experimental data. It successfully produces a shock-induced $\alpha\to\omega$ phase transition at similar pressures to those measured experimentally, and in fact produces Silcock- and TAO-I-oriented $\omega$ grains in approximately the correct proportion. As we have seen, it also predicts a diffuse diffraction signal with an azimuthal structure identical to that seen in experiment. However, the Zong potential also has shortcomings. On an `idealistic' level, the fact that it is neither entirely smooth nor continuous is obviously troubling. The consequences of the jumps in the potential energy surface are not catastrophic, but surely have a subtle influence on the system's dynamics that the user cannot easily quanity. There is a more `pragmatic' issue with the Zong potential, however, which is that it fails to reproduce one of the three $\alpha\to\omega$ transition mechanisms observed by Singh \textit{et al.}\ \cite{Singh2024}.

The atomistic pathway associated with this mechanism (which was not known at the time of the potential's creation, and, indeed, remains unknown at the time of writing) did not form part of the \textit{ab initio} energy database on which the potential was trained. It appears that the Zong potential (whose mathematical form is essentially aphysical) is unable to correctly extrapolate the potential energy surface into that unmapped region of configuration space. This is reflective of a more general observation made about machine-learning-class potentials by Nicholls \textit{et al.}\ \cite{Nicholls2023}, namely that they excel in the physical domain for which they are specifically trained, but often exhibit little to no transferability beyond this domain. Thus, there is a pressing need to identify the $\alpha\to\omega$ atomistic pathway associated with the third $\omega$ variant observed by Singh \textit{et al.}~\cite{Singh2024}, and to incorporate its associated configuration space into the training set of a new and physically robust potential for zirconium.

Finally, we note that much of the rich post-shock microstructure that the simulations predict -- including the residual, kinetically stabilized nanograins and interstitial defective material -- is contingent on the existence of the transient intermediate $\beta$-phase. Were the $\alpha\to\omega$ transition to occur directly, rather than via the two-stage Usikov-Zilberstein mechanism, we speculate that the resulting Hugoniot state might instead be a relatively `pure' $\omega$ nanocrystal that generates comparatively little non-thermal diffuse scattering. Experimental evidence for the intermediate $\beta$-phase in dynamically compressed Zr does indeed exist. Armstrong \textit{et al.}\ reported direct observation of metastable $\beta$-Zr using \textit{in situ} x-ray diffraction timed only hundreds of picoseconds after the onset of compression \cite{Armstrong2021}. In their nanosecond-duration experiments, Singh \textit{et al.}\ also showed that there existed weak diffraction peaks whose locations were consistent with a small phase fraction of residual $\beta$-Zr \cite{Singh2024}. For these reasons, we see no reason to doubt the predicted formation of the transient $\beta$-phase and all the attendant microstructure responsible for the non-thermal diffuse scattering.

%  5. Conclusion
\section{\label{sec:conclusion} Conclusion}

We have performed large-scale classical molecular dynamics simulations of [0001]-oriented single-crystal zirconium under dynamic uniaxial compression to 15~GPa using a machine-learning-class interatomic potential. The simulations predict a pressure-induced $\alpha\to\omega$ phase transition mediated primarily via a modified form of the displacive Usikov-Zilberstein mechanism, wherein $\omega$ grains heterogeneously nucleate at the twin boundaries of a short-lived intermediate $\beta$-phase. We further observe diffusive motion initiated at the junctions between $\beta$ grains, which leads ultimately to the formation highly defective $\omega$-like material at the interstices between $\omega$ grains.

We generated synthetic diffraction patterns from the simulated late-time atomistic configuration and compared its structure directly with that of femtosecond diffraction patterns acquired in dynamic-compression experiments. Both simulation and experiment revealed a sixfold-symmetric, azimuthally structured diffuse scattering signal beneath the conventional Bragg-like diffraction pattern. The simulations indicate that the anisotropic diffuse scattering stems from a confluence of thermal diffuse scattering, nanoparticle-like scattering from residual $\alpha$ and $\beta$ grains that survive the phase transition, and from the diffuse scattering components of the interstitial defective material. We demonstrated that the sparse nature of the monocrystalline target's reciprocal space is essential for the observation of the structured diffuse scattering.

These results demonstrate the interpretative power of atomistic simulation techniques in the field of dynamic compression, and underscore the importance of seeking direct, like-for-like comparisons between synthetic and experimental diffraction data. Diffuse scattering from dynamically loaded metals may yet prove to be a novel diagnostic of temperature and microstructure in which atomistic modeling techniques will play an essential role.

%  Acknowledgments
\section*{Acknowledgments}
The authors would like to express their gratitude to H.\ Zong and G.\ J.\ Ackland for sharing the interatomic potential used to execute the documented simulations. The authors also thank J.\ S.\ Wark, P.\ Svensson, and P.\ A.\ Burr for the insights they offered during the development of this work. This research was supported by the Laboratory Directed Research and Development (LDRD) Program at Lawrence Livermore National Laboratory (LLNL) (project Nos.\ 17-ERD-014 and 21-ERD-032). This work was performed under the auspices of the US Department of Energy by LLNL under Contract No.\ DE-AC52-07NA2734. P.\ G.\ H.\ gratefully acknowledges the support of AWE via the Oxford Centre for High Energy Density Science (OxCHEDS), and further support from EPSRC under Grant No.\ EP/X031624/1.

%  Bibliography
\bibliography{Bibliography_Main.bib}

\end{document}

% Title and authors
\title{Supplementary Material: Diffuse scattering from dynamically compressed single-crystal zirconium following the pressure-induced $\alpha\to\omega$ phase transition}
\author{P.~G.~Heighway}
%\email{patrick.heighway@physics.ox.ac.uk}
\affiliation{Department of Physics, Clarendon Laboratory, University of Oxford, Oxford, UK}
\author{S.~Singh}
\affiliation{Lawrence Livermore National Laboratory, Livermore, CA, USA}
\author{M.~G.~Gorman}
\affiliation{Lawrence Livermore National Laboratory, Livermore, CA, USA}
\author{D.~McGonegle}
\affiliation{AWE, Aldermaston, Reading, Berkshire, UK}
\author{J.~H.~Eggert}
\affiliation{Lawrence Livermore National Laboratory, Livermore, CA, USA}
\author{R.~F.~Smith}
\affiliation{Lawrence Livermore National Laboratory, Livermore, CA, USA}
\date{\today}
\date{\today}
\begin{abstract}
    This Supplementary Material details: the approximation of the Hugoniot predicted by the Zong potential~\cite{Zong2018} in the $\omega$ stability region; an illustration of the discontinuities in the Zong potential; a comparison of shock-MD simulations performed with and without the iterative `grow-and-prune' scheme; the fitting of the phase-transition-front velocity profile used to control the speed of the second confining momentum mirror; an estimation of the computational speedup offered by the grow-and-prune scheme; and a detailed description of the algorithm we used to separate atomistic configurations into their constituent phases.
\end{abstract}
\maketitle

%\section{Introduction}

%  S1. Calculation of Zong Hugoniot
\section{Approximation of the Hugoniot for the Zong potential}

The computational cost of the Zong potential makes a true mapping of its Hugoniot via a series of shock-compression simulation impracticable. We settle instead for an approximation based on solutions of the third Rankine-Hugoniot equation,
\begin{equation}\label{eq:RH3}
    E - E_0 = \frac{1}{2}P(V_0 - V)\ ,
\end{equation}
which connects the energy density $E$, specific volume $V$ and pressure $P$ in the shock state, given the ambient energy density $E_0$ and specific volume $V_0$. To implement this scheme in a molecular dynamics (MD) framework, one constructs an atomistic system whose microstructure is consistent with the expected post-shock state, and then systematically searches pressure-temperature space for states that satisfy Eq.~(\ref{eq:RH3}). For simplicity, we model the Hugoniot state above the $\alpha\to\omega$ phase-transition pressure as a defect-free, shear-stress-free monocrystal of $\omega$-Zr, and thus obtain only an approximation of the locus of shock states accessible under the Zong potential.

% Figure S1: Rankine-matching diagrams for calculation of Hugoniot
\begin{figure}[b]
\centering
\includegraphics{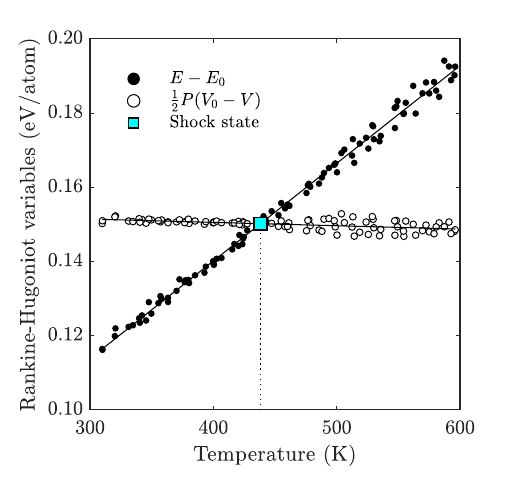}
\caption{\label{fig:rankine-matching} Example estimation of the temperature in a 12.5~GPa shock state for the Zong potential \cite{Zong2018} via numerical solution of the Rankine-Hugoniot equation $E - E_0 = \frac{1}{2}P(V_0 - V)$ (RH3), where $E$, $V$, and $P$ are the post-shock energy density, specific volume, and pressure, respectively, and $E_0$, $V_0$, and $P_0 = 0$ are their corresponding values for ambient $\alpha$-Zr. Simulations of perfect single crystals of $\omega$-Zr were conducted along a set of isobars between 5.0 and 20.0~GPa inclusive in steps of 2.5~GPa using a Nos\'{e}-Hoover baro-thermostat.}
\end{figure}

We first relax a perfect cubical crystal of fully periodic $\omega$-Zr containing 7\,000 atoms to 300~K and zero stress. We then increment the crystal's pressure by 5 GPa before propagating it along an isobar between temperatures of 300 and 600 K over the course of 10~ps using an $NPT$ ensemble. We control separate stress components independently, guaranteeing a hydrostatic stress state and allowing $\omega$-Zr to assume its most energetically favorable $a/c$ ratio at the given pressure. The energy density $E$ and specific volume $V$ are sampled every 0.1~ps. We sample isobars in this way for pressures between 5~GPa and 20~GPa in steps of 2.5~GPa.

After calculating the values of $E_0 = -6.19$~eV\,atom\textsuperscript{$-1$} and $V_0 = 23.6$ \AA\textsuperscript{3}\,atom\textsuperscript{$-1$} for ambient $\alpha$-Zr, we calculate the left- and right-hand sides of Eq.~(\ref{eq:RH3}) for each isobar, fit the data to linear functions, and then solve for their intersection, yielding the Hugoniot temperature at each pressure; this process is shown in Fig.~\ref{fig:rankine-matching}.

%  S2. Illustration of discontinuities in the Zong potential
\section{Discontinuities in the Zong potential}

\begin{figure}[t]
\centering
\includegraphics{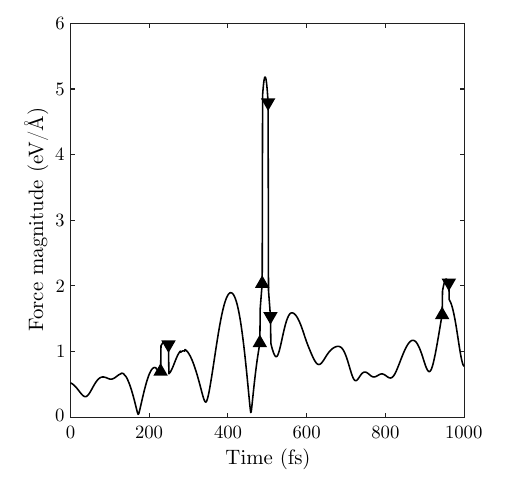}
\caption{\label{fig:discontinuities} Variation of the force experienced by an atom in a non-crystalline environment in Zr shock-compressed to 15~GPa. Arrows mark timesteps on which the the atom crosses discontinuities on the potential energy surface.}
\end{figure}

As has been previously shown by Nicholls \textit{et al.}~\cite{Nicholls2023}, the potential energy surface (PES) of the Zong potential possesses discontinuities. Sharp jumps in the PES (or rather its gradient) cause the forces experienced by atoms to sporadically change in a near-discontinuous manner. The time-history of the force felt by a representative atom is shown in Fig.~\ref{fig:discontinuities}. We observe that during a 1~ps interval, this atom experiences three brief ($\sim$10~fs) periods of abruptly amplified forces. The atom in question happened to be situated in a non-crystalline environment, but we have confirmed that atoms in the $\omega$-phase experience similar discontinuities.

% Figure S2: Example trajectory

As discussed in the Main Article, cumulative integration errors caused by the non-smoothness of the PES result in a slow drift of the system's energy and temperature (almost exactly 1 Kps\textsuperscript{$-1$}) even under nominal $NVE$ integration conditions. Due to the ill-conditioned nature of the potential, the temperature drift cannot practicably be stopped by reducing the simulation timestep; we found that reducing the timestep by an order of magnitude from 1.0~fs to 0.1~fs merely halved the drift rate. Instead, we combat the drift with a gentle global Nos\'{e}-Hoover thermostat that is just sufficient to keep the post-shock system at a temperature of 500~K.

%  S3. Comparison of simulations performed with and without grow-and-prune scheme
\section{Comparison of simulations performed with and without grow-and-prune technique}

It is important to verify that Stage I of the grow-and-prune technique (wherein an initially small crystal `bud' is incrementally grown in the shock propagation direction over the course of the simulation) yields the same results as a standard shock-MD simulation performed with a prebuilt, full-size crystal. To perform this comparison with the Zong potential itself would be extremely computationally demanding -- indeed, the very purpose of the grow-and-prune technique is to avoid having to simulate huge monolithic crystals. For that reason, we perform this comparison using a relatively inexpensive embedded-atom-method (EAM) potential for Zr designed by Mendelev and Ackland \cite{Mendelev2007}. Here, we will not focus on the details of the physics unfolding behind the shock; we seek only to compare the gross thermodynamic and microstructural evolution of crystals performed with and without the iterative construction scheme.

% Figure S2: Comparison of thermodynamic profiles obtained by 'Wallace' and 'Gromit' techniques
\begin{figure}[b]
\centering
\includegraphics{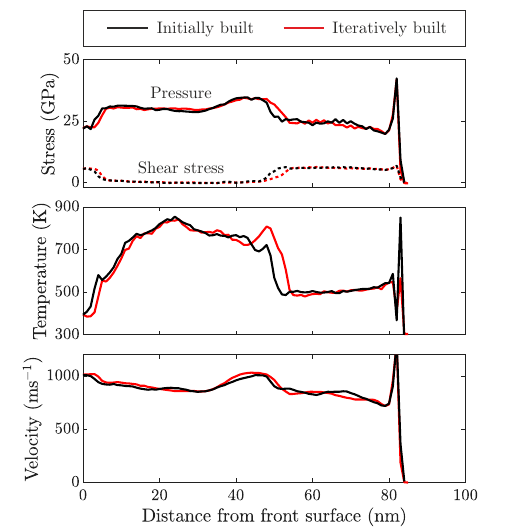}
\caption{\label{fig:wallace-gromit} Comparison of Eulerian stress, temperature, and particle-velocity profiles for a Zr single crystal shock-compressed along [0001], constructed either with its final dimensions from the outset (black) or iteratively (red). The shock was driven by a momentum mirror traveling at a fixed velocity of 1.0~kms\textsuperscript{$-1$}. Crystals were modeled under the relatively inexpensive Mendelev EAM potential \cite{Mendelev2007} for efficiency.}
\end{figure}

\begin{figure}[t]
\centering
\includegraphics{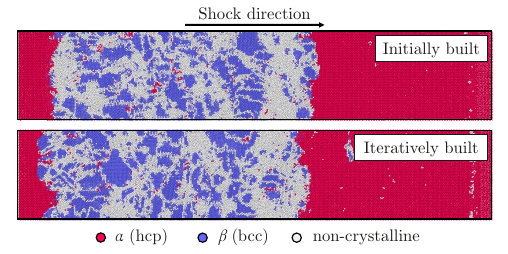}
\caption{\label{fig:comparison-of-cells} Visualization of the simulation cell for a Zr single crystal shock-compressed along [0001], constructed either with its final dimensions from the outset (above) or iteratively (below), with atoms colored by their local crystal structure using adaptive common-neighbor analysis (a-CNA) \cite{Stukowski2012b}. Corresponding thermodynamic profiles are shown in Fig.~\ref{fig:wallace-gromit}. Visualization performed using \textsc{ovito} \cite{Stukowski2010}.}
\end{figure}

We show in Fig.~\ref{fig:wallace-gromit} a comparison of the stress, temperature, and particle velocity profiles of a Mendelev-Zr single crystal shock-compressed with a momentum mirror traveling at 1.0 kms\textsuperscript{$-1$}, built either on the initial timestep or in an iterative manner, 17~ps into the simulation. We also provide visualizations of the two computational cells in Fig.~\ref{fig:comparison-of-cells} on the same simulation timestep. The dynamic response of the crystal under these conditions is relatively complex. After a purely elastic response lasting approximately 10~ps, we observe a pressure-induced $\alpha\to\beta$ phase transition initiated not nearest the drive surface, but at a finite depth into the crystal. The nascent bcc nanocrystal grows in both the positive and negative shock directions, and brings about markedly nonuniform thermodynamic conditions along the length of the crystal. Both Fig.~\ref{fig:wallace-gromit} and Fig.~\ref{fig:comparison-of-cells} show that the complex structure of the shock is essentially identical for the two simulation techniques; what differences exist are most likely attributable to differing initialization conditions. We are therefore confident that iterative construction of the crystal is a reasonable means of reducing the simulation cost without compromising the shock structure.

% Figure S3: Comparison of atomistic configuration obtained by 'Wallace' and 'Gromit' techniques

%  S3. Fitting of shock-front velocity
\section{Fitting of phase-transition-front velocity profile}

Stage II of the grow-and-prune technique involves deleting disposable material situated beyond the shock-compressed region of interest and keeping this region pressurized using a momentum mirror comoving with the new rear surface. To make the pruning step seamless, the velocity of the second mirror, $U_R$, should be forced to follow the local velocity evolution that would result if the disposable pressurizing material were still present. Since the material at which we perform the cut is (necessarily) close to the final Hugoniot state, we need only characterize the late-time, post-phase-transition portion of the particle-velocity evolution.

% Figure S3: Fitting of shock-front velocity profile
\begin{figure}[t]
\centering
\includegraphics{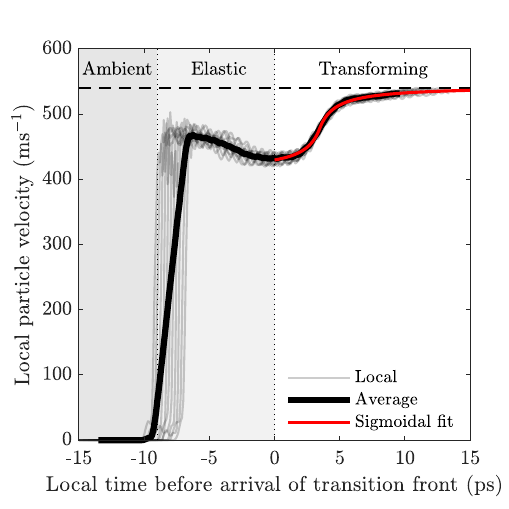}
\caption{\label{fig:relay-matching} Time-history of the local particle velocity for a set of eleven laminar Lagrangian material elements separated by 1~nm in a shock-compressed Zr crystal modeled using the Zong potential \cite{Zong2018}. For each element, time is defined such that the foot of the phase-transition front arrives at $t=0$~ps. The time-averaged velocity history is indicated by the thick black line. The velocity profile during the phase-transition period is fitted to an absolute sigmoid function.}
\end{figure}

Here, we approximate the particle velocity during and after the $\alpha\to\omega$ phase transition using a phenomenological absolute sigmoid function with a ceiling at $U_P$:
\begin{equation}\label{eq:mirror-velocity}
    u_p (t) = \min\left[ U_0 + \Delta U \left(\frac{\frac{t-t_0}{\Delta t_{\text{front}}}}{1 + \left|\frac{t - t_0}{\Delta t_{\text{front}}}\right|}\right), U_P \right]\ .
\end{equation}
We first obtain the shape of the sigmoidal part of the function (parametrised by variables $U_0$, $\Delta U$, and $\Delta t_{\text{front}}$) by calculating the average time-evolution of particle velocity and performing a fit. We calculate this average by extracting the local velocity profile of a series of material laminae and offsetting each by $z_L/U_S^p$ (where $z_L$ is their Lagrangian distance from the drive surface and $U_S^p$ is the Lagrangian velocity of the phase-transition front) to account for the time taken for the shock to reach each lamina. The collection of velocity profiles, their average, and the fitting to the sigmoidal function are shown in Fig.~\ref{fig:relay-matching}. The remaining parameter $t_0$ is then chosen such that the velocity of the second momentum mirror at the moment of its creation, $U_R(t_{\text{prune}})$, matches the local particle velocity at its initial position. The parameters used in our simulations are given in Tab.~\ref{tab:prune-parameters}.

% Table S1: Simulation parameters
\begin{table}[b]
\begin{tabular}{ccrl}
\hline
\hline
Variable                         & Symbol & \multicolumn{2}{c}{Value} \\
\hline
Particle velocity                & $U_P$ & 0.54 & kms\textsuperscript{-1} \\
Elastic precursor velocity       & $U_S^e$ & 5.4 & kms\textsuperscript{-1} \\
Phase-transition front velocity  & $U_S^p$ & 2.3 & kms\textsuperscript{-1} \\
Thermalization time              & $\Delta t_{\text{eq}}$ & 1.0 & ps \\
Pruning time                     & $t_I$ & 27.5 & ps \\
\multirow{4}{*}{Sigmoid fitting parameters} & $U_0$ & 4.77 & kms\textsuperscript{-1} \\
                                 & $\Delta U$ & 0.66 & kms\textsuperscript{-1} \\
                                 & $t_0$ & 22.2 & ps \\
                                 & $\Delta t_{\text{front}}$ & 1.26 & ps \\
\hline
\hline
\end{tabular}
\label{tab:prune-parameters}
\caption{Simulation parameters used in shock-compression simulations of single-crystal zirconium, using the `grow-and-prune' scheme. Sigmoid fitting parameters are those used to control the velocity of the second momentum mirror according to Eq.~(\ref{eq:mirror-velocity}).}
\end{table}

% Computational savings offered by grow-and-prune scheme
\section{Computational efficiency of grow-and-prune technique}

We estimate that for our Zr simulations, the grow-and-prune technique saves us a factor of approximately 8 in computational costs, versus a `naive' simulation wherein the crystal is built with its final dimensions from the outset. The general scaling of the computational efficiency with simulation duration can be understood as follows.

Suppose Stage I takes simulation time $t_I$. If the computational cost per unit length of crystal per unit simulation time is $c$, the total cost of Stage I is
\begin{equation}
    C_I = \int_0^{t_I} dt\,c \cdot (U_S^e t) = \frac{1}{2}cU_S^et_I^2\ ,
\end{equation}
where the precursor velocity $U_S^e$ dictates the rate at which the crystal grows with time.

If Stage II takes simulation time $t_II$, its computational cost is simply
\begin{equation}
    C_{II} = \int_0^{t_{II}} dt\,c \cdot (U_S^e t_I) = cU_S^et_It_{II}\ .
\end{equation}
\phantom{test}

If instead the crystal had been built with its final length $U_S^e (t_I + t_{II})$ from the outset, the cost of the simulation would have been
\begin{equation}
    C_0 = cU_S^e(t_I+t_{II})^2\ .
\end{equation}

By combining Eqs.~(S3-5), we can show that the ratio of the computational costs of the `naive' and grow-and-prune techniques, $R = C_0/(C_I+C_{II})$ is
\begin{equation}
    R = 2\left(1 + \frac{r^2}{1+2r}\right)\ ,
\end{equation}
where $r = t_{II}/t_{I}$ is the ratio of the durations of the two stages of the simulation. For our simulation, in which Stage I ends after 27.5~ps and Stage II lasts for a further 172.5~ps, we find $R \sim 8$. We therefore save ourselves almost an order of magnitude in computational costs using the grow-and-prune technique.

% Omega-identification technique
\section{Phase partitioning algorithm}

The pipeline we use to partition the simulated atomistic configurations into their constituent phases was executed using \textsc{ovito} \cite{Stukowski2010}. The algorithm comprises the following steps:
\begin{enumerate}
    \item Atoms are first classified according to their Ackland-Jones parameter (AJP) \cite{AcklandJones2006}. As originally shown by Zong \textit{et al.}\ \cite{Zong2019}, atoms in the $\omega$-phase take AJP values of either 2 or 3 (corresponding to nominal hcp and bcc structures, respectively) depending on which of the two inequivalent atomic environments they are located in; atoms situated on the basal-like planes assume AJP value 3, while those situated in the graphene-like planes between them take value 2. We will refer to these atoms as `basal' and `graphitic', respectively.
    \item We then identify atoms in the $\omega$-phase using the known coordination of the basal and graphitic atoms. To do so, we count the number of each type of atom ($n_b$ and $n_g$, respectively) within 3.5~\AA\ of the central atom. Within this radius, basal atoms should be coordinated with two basal atoms and 12 graphitic atoms, while graphitic atoms are coordinated with six basal and five graphitic atoms. Thus, basal and graphitic atoms whose coordination numbers $(n_b,n_g)$ take values of (2,12) or (6,5), respectively, are assigned to the $\omega$ phase. A tolerance of $\pm1$ is given to these coordination numbers to combat thermal fluctuations.
    \item Any atoms that are not assigned to the $\omega$-phase are then classified by adaptive common-neighbor analysis (a-CNA) \cite{Stukowski2012b}. This step identifies atoms in the $\alpha$- and $\beta$-phases. Any atoms not assigned to the $\alpha$-, $\beta$-, or $\omega$-phases is classified as non-crystalline.
    \item We then perform a final corrective step in which we identify and reassign `orphaned' atoms. An orphaned atom is one whose assigned structure differs from that of its neighbors, all of whom have been assigned the same structure. Each orphaned atom is reassigned to the structure taken by its neighbors. This step mitigates misidentifications caused by thermal fluctuations.
\end{enumerate}
Throughout the Main Article, the phase partitioning algorithm is applied to atomic coordinates time-averaged over a 200 fs window to further reduce thermal noise.

%  Bibiliography
\bibliography{Bibliography_Supplementary.bib}